# A Polarizable Water Potential Derived from a Model Electron Density

*Joshua A. Rackers,[1,2†] Roseane R. Silva,[1†] Zhi Wang[3] and Jay W. Ponder[3,4*]*

[1] Program in Computational & Molecular Biophysics, Washington University School of Medicine, Saint Louis, Missouri 63110, United States

[2] Center for Computing Research, Sandia National Laboratories, Albuquerque, New Mexico 87123, United States

[3] Department of Chemistry, Washington University in Saint Louis, Saint Louis, Missouri 63130, United States

[4] Department of Biochemistry & Molecular Biophysics, Washington University School of Medicine, Saint Louis, Missouri 63110, United States

† These authors contributed equally to this work.

* Corresponding author: ponder@dasher.wustl.edu

## Abstract

A new empirical potential for efficient, large scale molecular dynamics simulation of water is presented. The HIPPO (Hydrogen-like Intermolecular Polarizable POtential) force field is based upon the model electron density of a hydrogen-like atom. This framework is used to derive and parameterize individual terms describing charge penetration damped permanent electrostatics, damped polarization, charge transfer, anisotropic Pauli repulsion, and damped dispersion interactions. Initial parameter values were fit to Symmetry Adapted Perturbation Theory (SAPT) energy components for ten water dimer configurations, as well as the radial and angular dependence of the canonical dimer. The SAPT-based parameters were then systematically refined to extend the treatment to water bulk phases. The final HIPPO water model provides a balanced representation of a wide variety of properties of gas phase clusters, liquid water and ice polymorphs, across a range of temperatures and pressures. This water potential yields a rationalization of water structure, dynamics and thermodynamics explicitly correlated with an *ab initio* energy decomposition, while providing a level of accuracy comparable or superior to previous polarizable atomic multipole force fields. The HIPPO water model serves as a cornerstone around which similarly detailed physics-based models can be developed for additional molecular species.

## Introduction

Water is perhaps the most studied of all molecules, both experimentally and theoretically. In addition to its obvious importance for life on Earth, water is of interest due to: (1) its unique physical properties, including a density maximum near 4°C with normal ice being less dense than the liquid, (2) its ability to solvate a wide range of disparate chemical species, (3) the great variety of its solid-phase crystal forms and richness of its phase diagram, and (4) its paradigmatic hydrogen bonding interaction and the related hydrophobic effect. The first atom-based water potential available as a quantitative model dates back nearly a century to the work of Bernal and Fowler.[1] The ST2 model of Rahman and Stillinger,[2] among other models from that period, was suitable for use in some of the initial molecular dynamics (MD) simulations. During the early 1980s, the TIPS[3] and SPC[4] families of water potentials were developed, and they are still used in present day modeling projects. Since that time, a large number of additional water models have been proposed for use in simulation, ranging from coarse-grained empirical functions that represent several molecules by a single-site particle,[5, 6] to detailed density functional theory-based (DFT) MD calculations,[7] to massive simulations using machine-learned potentials.[8]

Here we propose a new water model near the intersection of empirical models fit to reproduce macroscopic properties, and *ab initio* models derived entirely from first-principles physics. This new model, referred to as HIPPO for Hydrogen-like Intermolecular Polarizable POtential, is derived directly from a model electron density obtained from *ab initio* results and electronic structure theory, but then parameterized to improve agreement with target experimental data. As such, the HIPPO water model provides a computationally efficient form for use in large-scale simulations, while allowing for analysis and decomposition in terms of physically validated energetic components.

In one sense, this new model is a natural extension of previous polarizable force fields. In particular, it extends the logic that has made the AMOEBA force field successful.[9, 10] The main advance of AMOEBA was to show that intermolecular interactions at medium range cannot always be handled through cancellation of errors, as they are in point charge force fields.[11] This insight motivated the inclusion of dipole polarization and atom-centered multipoles into the model. Much recent work, however, shows that despite AMOEBA's more elaborate functional form, it still relies on significant error cancellation at short range. The archetypal example of this behavior is the $\pi$-stacking interaction, exemplified by the benzene dimer.[12, 13] Studies of this system have shown that despite its atomic multipole and polarization terms, AMOEBA exhibits some of the same short-range problems as simpler force fields. A principal aim of the HIPPO model is to reduce this kind of reliance on error compensation at short range.

The way in which HIPPO achieves this aim, however, makes it more than a simple extension of AMOEBA. Much of the short-range error in force fields is due to their reliance on point approximations and the lack of an explicit charge density. In the $\pi$-stacking case, for example, the error in the electrostatic interaction is given a widely adopted name: charge penetration. Analogous errors occur in other force field components, but they all arise from the same inappropriate density treatment.

HIPPO addresses this problem directly by including a description of the electron density explicitly in the model. It is far from the first empirical potential to include a model for the density; other models, most notably the Gaussian Electrostatic Model (GEM),[14-17] have made use of explicit charge densities. However, HIPPO is the first force field to use an electron density model in constructing each component of the total potential. As we will detail in the Theory section, every non-valence term derives its form from charge densities and the interactions between them. This distinction makes HIPPO a new class of density-based model.

The choice of density-based form is not arbitrary, as HIPPO follows from Symmetry Adapted Perturbation Theory (SAPT) quantum energy decomposition analysis.[18, 19] SAPT divides the total interaction energy of a system into four physically meaningful components: electrostatics, polarization, Pauli repulsion and dispersion. Importantly and as its name implies, SAPT does this through the use of perturbation theory. The base, or unperturbed state, is represented by isolated molecules, and the energy components are computed as perturbations from that state as two molecules are allowed to interact. This perturbation theory logic lends itself well to classical approximation. As is detailed in previous work, each HIPPO term uses the atomic electron density model to construct a classical equivalent of the corresponding SAPT term.[20-22] In this way, HIPPO is not just parameterized against SAPT components; it can itself be considered a classical approximation of SAPT.

Conceptually, one might be tempted to assume the elaborate functional form of HIPPO would lead to a large increase in computational cost over similar polarizable models. This, however, misjudges the nature of the complexity in the underlying model. Atomic charge densities only overlap at short range, and the highest cost additions to HIPPO are restricted to the relatively few interactions in that regime. In this way, HIPPO is able to employ a more complex functional form while maintaining a computational cost roughly equal to that of other polarizable force fields.

The following sections provide: (1) a unified summary of the theoretical underpinnings of the portions the HIPPO force field needed for a water model, (2) a description of the computational and simulation methodologies used, (3) HIPPO results compared against quantum mechanical and experimental data for gas phase clusters, liquid water and ice, and (4) discussion of strengths and limitations of the HIPPO model and the suitability of SAPT as a framework for force field development.

## Theory

In the HIPPO force field, every atom is represented by two components: a model valence electron density and a core point charge. The atomic electron density, illustrated in Figure 1, emulates that of a hydrogen-like atom,

$$\rho_{HIPPO} = \frac{Q\zeta^3}{8\pi} e^{-\zeta r} + Z\ \delta(r) \tag{1}$$

where Q is the valence charge of the atom, Z is the core charge, $\zeta$ controls the width of the electron density, and $\delta$ is the Kronecker delta function. The HIPPO density also includes consistent higher-order atomic dipole and quadrupole terms for describing anisotropy. This model density is used to derive all four intermolecular energy terms that compose the HIPPO force field,

$$U_{HIPPO} = U_{electrostatic} + U_{induction} + U_{dispersion} + U_{Pauli\ repulsion}\ . \tag{2}$$

The general forms and derivations of these terms have been detailed in several references [20, 21, 22] describing the piecewise development of the model. To provide a unified picture, we present here a comprehensive definition of each term.

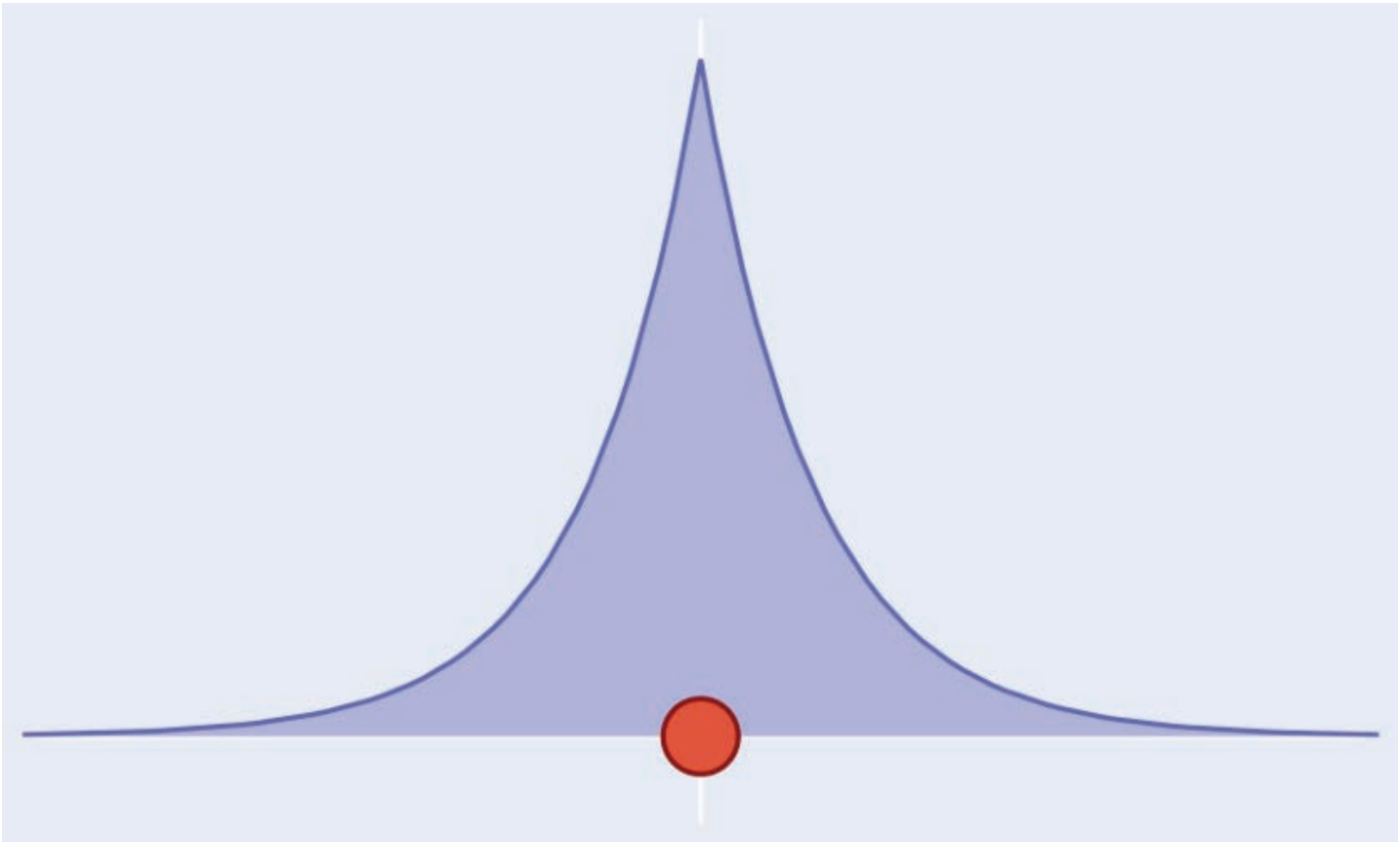

***Figure 1.*** *Schematic of a HIPPO atom. The blue shaded area represents the valence electron density, and the red point represents the point core charge.*

***Electrostatic Energy.*** Like its progenitor, the AMOEBA force field, the HIPPO electrostatic term is anisotropic, utilizing atomic multipole moments through the quadrupole. Since each atom in the model is represented by a core charge and a smeared density, the pairwise Coulomb interaction has four components. The HIPPO electrostatic energy is defined as,

$$U_{electrostatic}^{HIPPO} = \sum_{i>j} Z_i T_{ij} Z_j + Z_i \boldsymbol{T}_{\boldsymbol{ij}}^{*} \vec{M}_j + Z_j \boldsymbol{T}_{\boldsymbol{ji}}^{*} \vec{M}_i + \vec{M}_i \boldsymbol{T}_{\boldsymbol{ij}}^{\boldsymbol{overlap}} \vec{M}_j \tag{3a}$$

$$\vec{M} = \left( Q, [\mu_x, \mu_y, \mu_z], \begin{bmatrix} \Theta_{xx} & \Theta_{xy} & \Theta_{xz} \\ \Theta_{yx} & \Theta_{yy} & \Theta_{yz} \\ \Theta_{zx} & \Theta_{zy} & \Theta_{zz} \end{bmatrix} \right) \tag{3b}$$

$$T_{ij} = \frac{1}{r_{ij}} \tag{3c}$$

$$\boldsymbol{T}_{\boldsymbol{ij}}^{*} = [1 \quad \nabla \quad \nabla^2] \left( \frac{1}{r_{ij}} f_{ij}^{damp}(r_{ij}) \right) \tag{3d}$$

$$\boldsymbol{T}_{\boldsymbol{ij}}^{\boldsymbol{overlap}} = \begin{bmatrix} 1 & \nabla & \nabla^2 \\ \nabla & \nabla^2 & \nabla^3 \\ \nabla^2 & \nabla^3 & \nabla^4 \end{bmatrix} \left( \frac{1}{r_{ij}} f_{ij}^{overlap}(r_{ij}) \right) \tag{3e}$$

where the first term represents the core-core repulsion, the second and third terms represent the core–density attractions and the fourth term represents the density–density repulsion. The M vector contains the multipole moments (charge, dipole and traceless quadrupole) and Q and Z represent the core and density charges constrained to satisfy the relation for the total atomic partial charge $q_i = Z_i + Q_i$. The $f^{damp}$ and $f^{overlap}$ terms in equations 3d and 3e are of critical importance. They result directly from the electrostatic potential generated by the model density,

$$V(r) = \frac{Q}{r}\left[1 - \left(1 + \frac{1}{2}\zeta r\right)e^{-\zeta r}\right] \ . \tag{4}$$

This gives the core–density attractions,

$$U_{core-density} = Z_i V_j(r_{ij}) = Z_i\left(\frac{1}{r_{ij}} f_{ij}^{damp}(r_{ij})\right) q_j \tag{5a}$$

$$f_{ij}^{damp}(r_{ij}) = 1 - \left(1 + \frac{1}{2}\zeta_j r_{ij}\right)e^{-\zeta_j r_{ij}} \tag{5b}$$

yielding the "one-center" damping factor that goes into $\boldsymbol{T}^*$. The density–density repulsion is given by

$$U_{density-density} = \frac{1}{2}\left[\int \rho_i(\boldsymbol{r})V_j(\boldsymbol{r})dv + \int \rho_j(\boldsymbol{r})V_i(\boldsymbol{r})dv\right] = q_i\left(\frac{1}{r_{ij}} f_{ij}^{overlap}(r_{ij})\right) q_j \tag{6a}$$

$$f_{ij}^{overlap}(r_{ij}) = \begin{cases} 1 - \left(1 + \frac{11}{16}\zeta r_{ij} + \frac{3}{16}(\zeta r_{ij})^2 + \frac{1}{48}(\zeta r_{ij})^3\right)e^{-\zeta r_{ij}}, & \zeta_i = \zeta_j \\ 1 - A^2\left(1 + 2B + \frac{\zeta_i}{2} r_{ij}\right)e^{-\zeta_i r_{ij}} - B^2\left(1 + 2A + \frac{\zeta_j}{2} r_{ij}\right)e^{-\zeta_j r_{ij}}, & \zeta_i \neq \zeta_j \end{cases} \tag{6b}$$

$$\text{with } B = \frac{\zeta_i^2}{\zeta_i^2 - \zeta_j^2} \ \text{ and } \ A = \frac{\zeta_j^2}{\zeta_j^2 - \zeta_i^2} \ , \tag{6c}$$

where the integrals are evaluated according to the method of Coulson.[23] The $f^{overlap}$ term is the "two-center" damping factor necessary to compute the fourth term of the HIPPO electrostatic potential energy. The terms necessary for higher-order multipole interactions are obtained by successive gradient operations applied to each of the damping factors as specified in equations 3d and 3e. In the interest of clarity, the explicit equations for all orders of the multipole interaction energy are enumerated in Appendix A. In the limit of large $\alpha$, both damping factors tend to unity and the undamped point multipole interaction energy is recovered. In practice, the use of finite densities remedies the well-documented charge penetration problem of electrostatics.[13, 24-29] In total, the HIPPO electrostatic model has five parameters per atom: a core charge Z, a valence charge Q, a dipole moment $\mu$, a quadrupole moment $\Theta$, and a "charge penetration" damping parameter $\zeta$.

***Induction Energy.*** In addition to the permanent core charge and density-based multipoles, HIPPO includes a point inducible dipole at every atomic site. The induction energy of the model is defined as,

$$U_{induction}^{HIPPO} = \sum_{i} \frac{1}{2} \vec{\mu}_i^{ind} \vec{F}_i^{perm} - \sum_{i>j} \varepsilon_i e^{-\eta_j r_{ij}} + \varepsilon_j e^{-\eta_i r_{ij}} \tag{7}$$

where the first term represents the polarization energy of the induced dipoles interacting with the permanent electric field and the second term represents a small pairwise exponential charge transfer function. The polarization term is the source of many-body energy in the force field. The induced dipoles are determined by solving the system of linear equations,

$$\boldsymbol{\mu} = \boldsymbol{\alpha}\left(\boldsymbol{F^{perm}} + \boldsymbol{F^{ind}}\right) \tag{8}$$

where the vectors are defined as $\boldsymbol{\mu} = [\mu_1, \mu_2, \mu_3, \ldots, \mu_n]$ and similarly for $\boldsymbol{F}^{perm}$ (the field due to the permanent multipoles), $\boldsymbol{F}^{ind}$ (the field due to the induced dipoles) and $\boldsymbol{\alpha}$ (the atomic polarizabilities). The permanent and induced electric fields are calculated in the same manner, with the same parameters, as described in the previous section. In this way, the electric fields for the polarization model are completely consistent with the permanent electrostatics portion of the model. For completeness, the full equations describing polarization are detailed in Appendix B. The only additional parameter necessary for the polarization model is the atomic polarizability of each atom, denoted by $\alpha$. Finally, the charge transfer function requires two parameters per atom: a prefactor $\varepsilon$ and an atom-based damping factor $\eta$.[30]

***Dispersion.*** The dispersion interaction between atoms arises from the interaction energy of correlated, instantaneous induced dipole moments. In the point approximation, this gives the canonical $1/r^6$ dependence associated with London dispersion.[31] Because the HIPPO model represents valence electrons as densities, the functional dependence is somewhat modified. The dispersion energy between two atoms with instantaneous induced dipoles, $\mu_i$ and $\mu_j$, is found by solving Schrödinger's equation,

$$\frac{1}{M}\frac{\delta^2\Psi}{\delta z_i^2} + \frac{1}{M}\frac{\delta^2\Psi}{\delta z_j^2} + \frac{2}{\hbar}\left(E - \frac{1}{2}kz_i^2 - \frac{1}{2}kz_j^2 - U_{dipole-dipole}\right)\Psi = 0 \tag{9}$$

where, for the case of correlated, parallel dipoles,

$$U_{dipole-dipole} = \mu_i\left(\nabla^2\left(\frac{1}{r_{ij}}f_{ij}^{overlap}(r_{ij})\right)\right)\mu_j = \frac{\mu_i\mu_j}{r^3}\lambda_3^{overlap} - \frac{3(\mu_i r)(\mu_j r)}{r^5}\lambda_5^{overlap}$$
$$= \frac{\mu_i\mu_j}{r^3}\left(3\lambda_5^{overlap} - \lambda_3^{overlap}\right) = \frac{\mu_i\mu_j}{r^3}f_{damp}^{dispersion} \; . \tag{10}$$

The damping factors, $\lambda_3$ and $\lambda_5$, that define $f_{damp}$ for dispersion are derived from the action of the gradient operator and are identical to those for the dipole-dipole interaction energy as defined in Appendix A. Solving the Schrödinger equation from equation 9 yields,

$$E = \frac{1}{2}\hbar(\omega_1 + \omega_2) \tag{11}$$

$$\omega_1 = \omega_0\sqrt{1 - \frac{2Q^2}{r^3 k} f_{damp}^{dispersion}}\ , \quad \omega_2 = \omega_0\sqrt{1 + \frac{2Q^2}{r^3 k} f_{damp}^{dispersion}}\ . \tag{12}$$

This energy expression can be effectively approximated with a binomial expansion,

$$\sqrt{1+x} = 1 + \frac{1}{2}\mathrm{x} - \frac{1}{8}\mathrm{x}^2 + \cdots \tag{13}$$

and the total energy thus becomes,

$$E = \hbar\omega_0 - \frac{Q^4\hbar\omega_0}{2r^6k^2} + \cdots \ . \tag{14}$$

Subtracting the energy of two infinitely separated dipoles ($\hbar\omega_0$) and substituting the parameter $C_6$ for $\frac{Q^2\sqrt{\hbar\omega_0}}{\sqrt{2}k}$ gives the pairwise dispersion energy,

$$U_{dispersion}^{HIPPO} = -\sum_{i<j} \frac{C_6^i C_6^j}{r^6} \left(f_{damp}^{dispersion}\right)_{\mathrm{ij}}^2 \ . \tag{15}$$

It is well known that accurate modeling of the dispersion energy at short range requires the use of a damping function.[32-41] HIPPO provides a non-empirical damping function derived from the dipole density-dipole density interaction. The model requires only one $C_6$ parameter per atom since the parameters for the damping function are fixed at their electrostatic model values.

***Pauli Repulsion.*** The final element of the HIPPO model is a density-based, multipolar model for Pauli Repulsion. Pauli repulsion is a consequence of the rearrangement of electron density that occurs when the Pauli exclusion principle is applied to electron densities of two unperturbed interacting molecules.[42-49] In previous work, we show that the primary change in electron density, relative to the unperturbed reference state, is an evacuation of electron density from the internuclear region.[22] The energy associated with this accumulation of charge in the internuclear region is proportional to

$$U_{Pauli\ repulsion} \propto \frac{S^2}{R} \tag{16a}$$

$$S = \int \phi_i \phi_j dv \tag{16b}$$

where $S$ is the overlap integral between the atomic orbitals on $i$ and $j$, and $R$ is the internuclear distance. To obtain suitable quantities to implement this model, we use the ansatz

$$\rho = \phi^* \phi \tag{17}$$

to define real, atomic pseudo-orbitals as:

$$\phi = \sqrt{\rho} = \sqrt{\frac{Q\zeta^3}{8\pi}} e^{\frac{-\zeta r}{2}} \ . \tag{18}$$

These pseudo-orbitals define the charge-charge portion of the overlap integral,

$$S = \int \phi_i \phi_j dv = \sqrt{Q_i Q_j \zeta_i^3 \zeta_j^3} \left[ \frac{1}{2X^3 R} \left( \zeta_i \left( RX - 2\zeta_j \right) e^{\frac{-\zeta_j R}{2}} + \zeta_j (RX + 2\zeta_i) e^{\frac{-\zeta_i R}{2}} \right) \right] \tag{19}$$

with

$$X = \left(\frac{\zeta_i}{2}\right)^2 - \left(\frac{\zeta_j}{2}\right)^2 \ .$$

From the bracket term, we can define

$$f_{exp}^{repulsion}(R) = \begin{cases} \frac{1}{\zeta^3}\left(1+\frac{\zeta R}{2}+\frac{1}{3}\left(\frac{\zeta R}{2}\right)^2\right)e^{\frac{-\zeta R}{2}}, & \zeta_i = \zeta_j \\ \frac{1}{2X^3R}\left[\zeta_i(RX-2\zeta_j)e^{\frac{-\zeta_j R}{2}} + \zeta_j(RX+2\zeta_i)e^{\frac{-\zeta_i R}{2}}\right], & \zeta_i \neq \zeta_j \end{cases}. \quad (20)$$

This allows writing $S^2$ in the familiar Coulombic form,

$$\frac{S_{charge-charge}^2}{R} = Q_i T_{pauli} Q_j \quad (21)$$

with

$$T_{pauli} = \frac{\zeta_i^3 \zeta_j^3}{R}\left(f_{exp}^{repulsion}\right)^2 \quad (22)$$

where $T_{pauli}$ (and, in turn, $S^2$) is dominated at short range by the exponential $f^{repulsion}$ term.

The anisotropy of the HIPPO repulsion model is obtained through its use of atomic multipole moments. Because $S^2$ has a clearly Coulombic form, we can include higher-order terms in the same manner as for electrostatics,

$$\frac{S_{total}^2}{R} = \sum_{i>j} \vec{M}_i \boldsymbol{T_{ij}^{repulsion}} \vec{M}_j \quad (23a)$$

where

$$\vec{M} = \left(Q, [\mu_x, \mu_y, \mu_z], \begin{bmatrix} \Theta_{xx} & \Theta_{xy} & \Theta_{xz} \\ \Theta_{yx} & \Theta_{yy} & \Theta_{yz} \\ \Theta_{zx} & \Theta_{zy} & \Theta_{zz} \end{bmatrix}\right) \quad (23b)$$

$$\boldsymbol{T_{ij}^{repulsion}} = \begin{bmatrix} 1 & \nabla & \nabla^2 \\ \nabla & \nabla^2 & \nabla^3 \\ \nabla^2 & \nabla^3 & \nabla^4 \end{bmatrix}\left(T_{pauli}\right). \quad (23c)$$

The multipole moments used are identical to those from the electrostatics calculation and $T^{repulsion}$ is a natural generalization of $T_{pauli}$. The interpretation here is that just as the charge component of the multipole expansion has a density, so too do the dipole and quadrupole moments. The various multipolar terms described in equation 23 represent the overlaps between the pseudo-orbitals associated with each individual density component. This definition of $S^2$ allows us to establish an anisotropic repulsion model we call the Multipolar Pauli Repulsion model,

$$U_{Pauli\ repulsion}^{HIPPO} = \sum_{i<j} \frac{K_i K_j}{r_{ij}} S_{total}^2 \quad . \tag{24}$$

A complete derivation of this model is detailed in our previous work.[22] Full equations defining the model as presented here, with higher-order terms included, are presented in Appendix C. The HIPPO repulsion model introduces three parameters per atom: a proportionality constant K, an exponential parameter $\alpha$, and a valence charge Q. Note that although analogous to their counterparts in the electrostatics derivation, the parameters $\zeta$ and Q are allowed to differ from their adopted values in the electrostatic energy term.

***Valence Terms.*** The HIPPO water model is fully flexible. It includes a bond stretching term and angle bending term, whose functional forms are the same modified harmonic potentials used in AMOEBA[9] and originally taken from work by Allinger on the MM3 force field.[50] Stretch-bend and Urey-Bradley coupling terms are not used. HIPPO does include a charge flux term which couples the atomic partial charges with the H-O stretching motions and the H-O-H angle,[51] and serves to provide a dipole moment derivative surface in better agreement with quantum mechanical calculations.[52] Previous work with the AMOEBA+ force field has shown that this charge flux term correctly reproduces the average increase in the H-O-H angle, from 104.5$^o$ to roughly 106$^o$, that occurs when transferring water from gas to liquid phase.[53] The inclusion of this term, with parameters optimized for the HIPPO water model, yields the same correct behavior for the average angle value.

## Methods

***Code Implementation.*** HIPPO calculations in this paper were performed with the Tinker Version 8, Tinker-OpenMM, and Tinker9 packages.[54-56] Implementation of HIPPO was undertaken by Josh Rackers and Jay Ponder in Tinker, Joshua Rackers, Zhi Wang and Roseane Silva in Tinker-OpenMM, and Zhi Wang and Roseane Silva in Tinker9. Molecular dynamics simulations data in the paper were performed with Tinker9 on our in-house GPU cluster. All subsequent analysis was performed using Tinker on workstation CPU hardware and Tinker9 on the GPU cluster.

The Tinker9 code is optimized for standard simple partial charge force fields and for the AMOEBA potential, while the HIPPO code is unoptimized. Molecular dynamics benchmarks for three 24051 atoms, 62.23 Å cubic water boxes using current Tinker9 code and an NVIDIA 3070 Ti GPU are as follows: TIP3P, 325.5 ns/day (2.0 fs steps, rigid water via SETTLE); AMOEBA, 29.1 ns/day, and HIPPO 24.6 ns/day (both run with 2.0 fs steps, RESPA multiple time step integrator, SCF induced dipole convergence to 0.00001 Debye RMS). A looser induce dipole convergence of 0.01 D is sufficient for many production calculations, and its use increases the speed of AMOEBA and HIPPO to 43.4 ns/day and 33.6 ns/day, respectively. Based on comparative timings with CPU code, we estimate that fully optimized Tinker9 HIPPO code will be at least as fast as AMOEBA, and likely about 25% faster.

In order to facilitate model development, our current HIPPO implementation is written for ease of modification instead of for computational speed. First, multipole, polarization and repulsion terms are computed in independent, modular code sections, requiring redundant evaluation of the geometric and interaction terms for dipoles and quadrupoles. Second, the multipole and polarization are directly computed in the global Cartesian coordinate frame, without use of spherical harmonics or prior rotation of pair interactions into quasi-internal frames.[57] Speed advantages for HIPPO compared to AMOEBA include the use of particle mesh Ewald summation (PME) for dispersion interactions,[21] and HIPPO's simpler gradient computation due to its use of unified exclusion and scaling rules for induced dipole and energy calculations.

***Parameterization Procedure.***

*Stage One: Fit to SAPT data*

The initial multipole and valence parameters were fit to monomer data. The multipole parameters were obtained using a protocol analogous to that for AMOEBA parameterization[10] and initial bond and angle parameters were taken from AMOEBA. The rest of the initial parameters pertaining to the intermolecular potential were fit exclusively to SAPT data. SAPT2+ reference calculations with an aug-cc-pVDZ basis set were performed on 27 water dimer structures. These structures included seven points on the dissociation curve, ten points on the canonical dimer angular surface, and the ten stationary point dimer structures of Smith, *et al*.[58] Each term of the force field was fit to its corresponding component from the SAPT decomposition. Observations on the quality of the resulting parameters can be found in the discussion section.

*Stage Two: Constrained Genetic Algorithm Search*

The initial parameter set obtained through fitting to SAPT energy components needed further adjustment to better match condensed phase properties. To improve the liquid water properties while keeping the features of the SAPT fitting, we continued optimizing the model by performing a global search in parameter space centered at the initial values. A differential evolution optimizer from the Scipy 1.8 package was used. The objective function of this optimizer has two main components: the energy decomposition of the Smith dimers and the heat of vaporization of water at room temperature.

While this optimizer was generating liquid data, a second function was simultaneously evaluating the liquid properties from each simulation. This function was merely a tool to select simulations with desired properties. The goal was to find simulations with liquid density, heat of vaporization and self-diffusion coefficient within 1% of their experimental values. The search was ended upon generation of five parameter sets satisfying all requirements. One of these sets was chosen as the best to continue the parametrization.

*Stage Three: Parameter Refinement with ForceBalance*

Following the global search in parameter space, we used a least square optimizer to fit a wider range of properties and to guarantee we were at a local minimum in parameter space. For this step, we used the ForceBalance (FB) program.[59] The goal of this final parameterization step was to obtain a model with desired condensed phase properties across a wide range of temperature and pressure.

The distinctive feature of FB is its ability to compute parametric derivatives of condensed phase properties from MD simulations using thermodynamic fluctuation equations. To refine parameters, we set a minimal objective function including experimental densities, enthalpies of vaporization, and dielectric constants over a range of temperatures from 261 K to 373 K. No other condensed phase properties were considered in the fitting procedure.

***Computational Details.*** All properties and simulations were obtained using the HIPPO force field as implemented in the Tinker and Tinker9 packages. To compute condensed phase properties, MD simulations of liquid water were performed. Unless otherwise noticed, properties were computed based on simulation of a cubic box of dimension ~50 Å and containing 4,200 water molecules. The thermodynamics properties listed in Table 4 were calculated from simulations at constant pressure and temperature. All simulations were performed using the RESPA (Reversible Reference System Propagator Algorithm) integrator coupled with a Monte Carlo barostat[60] and the Bussi thermostat.[61] For FB fitting, each MD simulation ran for 2 ns using a 2.0 fs time step, with a 0.5 ns equilibration phase and 1.5 ns production phase. The energy components of water dimers and clusters were calculated using the ANALYZE program in Tinker.

The temperature dependence of water properties was computed from a total of 40 simulations carried out at atmospheric pressure (1 atm), for temperatures ranging from 248 K to 373 K. Each simulation was started at the experimental density for the respective temperature and ran for at least 20 ns using a 2.0 fs time step; the first 2 ns of the simulations were discarded as equilibration. For temperatures less than 300 K, the production MD was extended by 10 ns to guarantee convergence of properties.

The self-diffusion coefficient was computed following the steps described for the MB-Pol model.[62] We chose to run simulations in ~100 Å cubic boxes with 33,500 water molecules. The larger box size was used to reduce known finite size effects in the calculation of self-diffusion coefficients.[63] We simulated 26 temperatures in total, ranging from 248 K to 373 K. For each simulation temperature, a box was built such that its density matched the experimental density for that temperature. Then, each simulation box was equilibrated for 0.5 ns in an NPT ensemble at atmospheric pressure. Following equilibration, we ran an additional 1.5 ns trajectory. From this trajectory, thirty different structures were selected, at 50 ns intervals. From those structures, thirty independent NVT trajectories of 100 ps were obtained. Then, we ran 100 ps simulations in an NVE ensemble. The self-diffusion coefficient was computed from each NVE trajectory and averaged over the 30 independent calculations for each temperature.

In order to evaluate finite size effects in computation of the self-diffusion coefficient, we ran additional simulations with different box sizes at room temperature (298 K). Each simulation was run for 4 ns in NVT ensemble, and the self-diffusion coefficient was computed using the final 3.5 ns of data. Five cubic box simulations were performed: 300 water molecules in ~20 Å box, 900 molecules in ~30 Å box, 4,200 molecules in a ~50 Å box, 17,100 molecules in a ~80 Å box, and finally 33,500 molecules in a ~100 Å box.

To calculate the surface tension of liquid water, we first selected four starting structures, at least 100 ps apart, from the production phase of our NPT simulations at different temperatures. Each structure was then simulated for 500 ps in an NVT ensemble. Then the Z-axis of each cubic box was expanded to three times the X-axis and Y-axis dimensions.[64] The final system geometries were slabs with X = Y = ~ 50 Å, and Z = ~150 Å, with a vacuum layer along the Z-axis over each side of the slab. Each system was then simulated in the NVT ensemble for 10 ns. The surface tension was calculated from the last 9.5 ns of data using the pressure tensor,[65] which was computed every picosecond. The final surface tension value reported for each temperature is the average of the four independent calculations.

The pressure dependence of the liquid water density was computed from a total of 10 simulations at room temperature (298 K), and with target pressure ranging from 1 atm to 9000 atm. Each simulation was started at the experimental density for the respective pressure and run for 10 ns using a 2.0 fs time step, with the first 2 ns as equilibration. The cubic box size for this set of simulations was ~30 Å with 900 water molecules.

We selected eight ice crystal structures to compute lattice energy and density. Ice energies were computed after energy minimization of the initial structure using a steepest descent algorithm. Each minimized structure was then simulated for 10 ns in the NPT ensemble. The average density of each ice crystal was computed using the last 8 ns of trajectory data. The target temperature and pressure for each simulation were set to the respective values reported for each polymorph crystal structure.

## Results

Because both the functional form and parameterization of the HIPPO water model are rooted in quantum mechanics, we set out to test the accuracy of the model against both experimental condensed phase data and *ab initio* calculations. In this section we will move from small to large clusters, starting from the properties of the water monomer and dimers, through successively larger clusters and up to condensed phase. By showing this behavior across scales we hope to demonstrate the power of a first-principles derived potential energy function.

***Parameters.*** Full specification of the HIPPO force field water model includes 37 refined parameter values. Explicit values for these parameters with their associated units are provided in the Supporting Information as Table S1. Several of the parameters are highly correlated, such that the effective number of parameter degrees of freedom required for the HIPPO model is lower than the number of raw parameters. While many of the parameter classes also used in previous AMOEBA-like water models, such as the atomic multipole values on oxygen and hydrogen, adopt similar values in HIPPO, the differences observed are important to the accurate reproduction of many water properties. Finally, where earlier work on individual components of the HIPPO model considered additional molecules,[20-22] the HIPPO water values reported here are in line with periodic trends across these other molecules and atom types.

***Monomer.*** The foundation of the HIPPO model is the monomer electron density. The fidelity of the rest of the model relies on an accurate representation of the true electron density of the molecule. Table 1 shows that HIPPO reproduces the monomer multipole moments and polarizability of an isolated water molecule with a satisfactory level of agreement.

***Table 1.*** *HIPPO Water Monomer Properties. All calculations performed on experimental, gas phase geometry where the Z-axis is the $C_2$ axis, the molecule lies in the XZ-plane, and the O atom is along the negative Z-axis.*

| | Dipole (D) | Quadrupole (B) | | | Polarizability (Å$^{-3}$) | | |
|---|---|---|---|---|---|---|---|
| | $d_z$ | $Q_{xx}$ | $Q_{yy}$ | $Q_{zz}$ | $a_{xx}$ | $a_{yy}$ | $a_{zz}$ |
| **HIPPO** | 1.843 | 2.48 | -2.38 | -0.10 | 1.613 | 1.289 | 1.362 |
| **Experiment** | 1.855 | 2.63 | -2.50 | -0.13 | 1.528 | 1.415 | 1.468 |

Additionally, HIPPO accurately reproduces the electrostatic potential around the water monomer as illustrated in Figure 2. The “Multipole Only” panel shows the signature of the “charge penetration” effect with a large negative error near the molecular surface. The point multipole model systematically underestimates the electrostatic potential at short range. Previous work has shown that including a simple density model can largely eliminate this charge penetration error, and this is clearly true for the HIPPO model. The "HIPPO" panel in Figure 2 shows that error in the electrostatic potential at short range is greatly reduced relative to the undamped point multipole model.

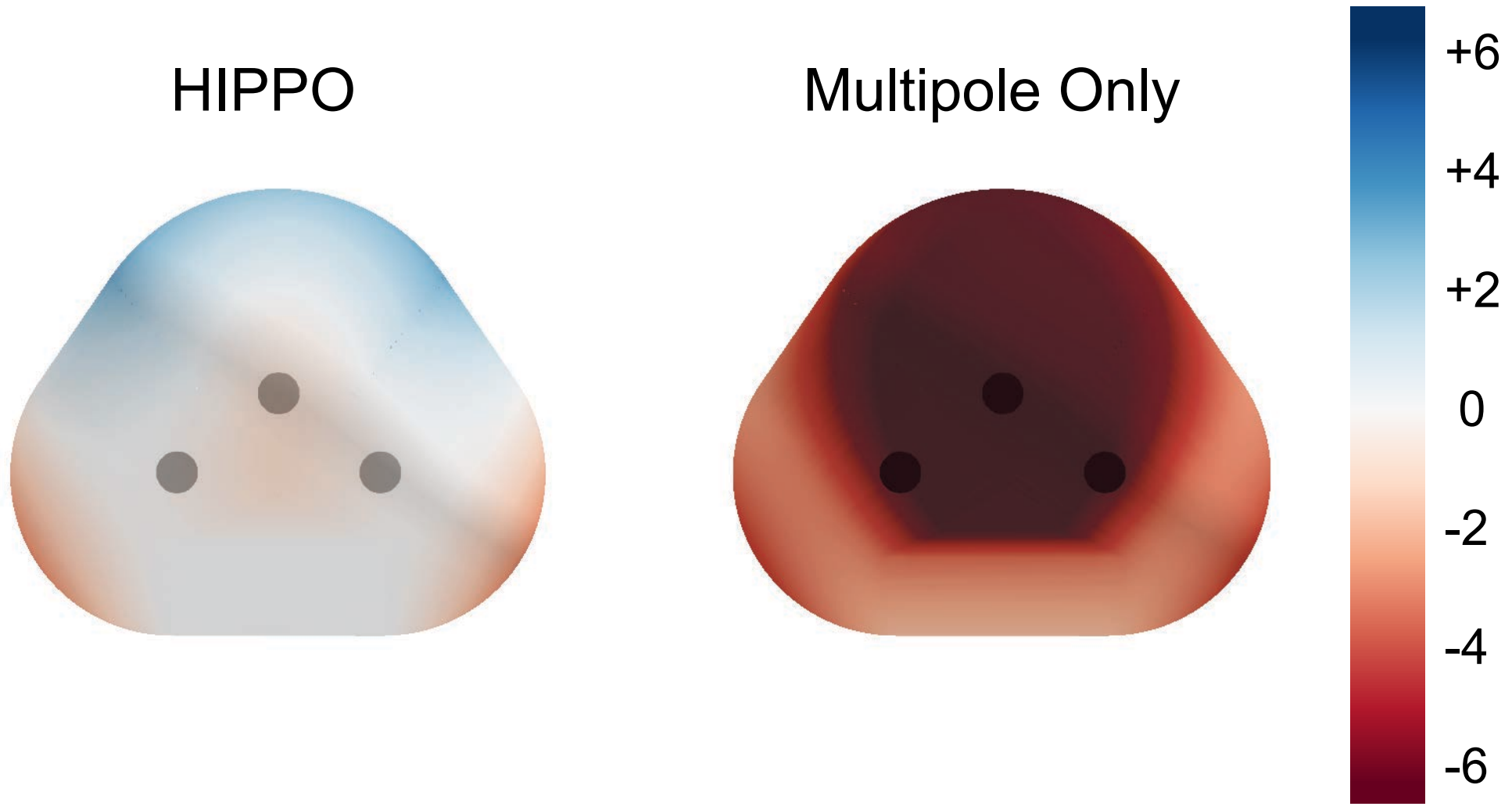

***Figure 2.*** *Error in the electrostatic potential: HIPPO vs. point multipoles neglecting charge penetration. The plot on the right shows the error in the electrostatic potential at the van der Waals surface for the undamped point multipole model. The plot on the right shows the same for HIPPO. Both use the same set of multipoles through quadrupole. Values are given in kcal/mol/electron.*

***Dimers.*** The water dimer potential energy surface is foundational to the overall model because it is the first place where the entire intermolecular energy function comes into play. For HIPPO in particular, this surface is of tremendous importance as the density-based terms of the intermolecular potential energy function are constructed specifically to reproduce dimer intermolecular interactions. Because it has been extensively studied, we have selected three separate "slices" of the dimer potential energy surface on which to evaluate the HIPPO model: the canonical water dimer dissociation curve, the angular dependence of the water dimer hydrogen bond angle, and the ten well-studied stationary points of Smith *et al.*[58] For each of these slices we evaluate the HIPPO model relative to two references. First, we compare the total energies of HIPPO to the total energies from *ab initio* calculations. Second, we compare the components of the HIPPO intermolecular potential energy function to their corresponding components from a SAPT decomposition.

The dissociation curve of the canonical water dimer is an important piece of the dimer potential energy surface because it contains information about the balance between short-range effects like repulsion and charge penetration, and long-range effects such as dispersion and multipole electrostatics. To generate this curve, we took the water-water interaction structures from the S101x7 database.[29] These structures represent the water dimer at points from 0.7 to 1.1 times the equilibrium distance. The results for HIPPO *vs*. the *ab initio* reference data are plotted in Figure 3A.

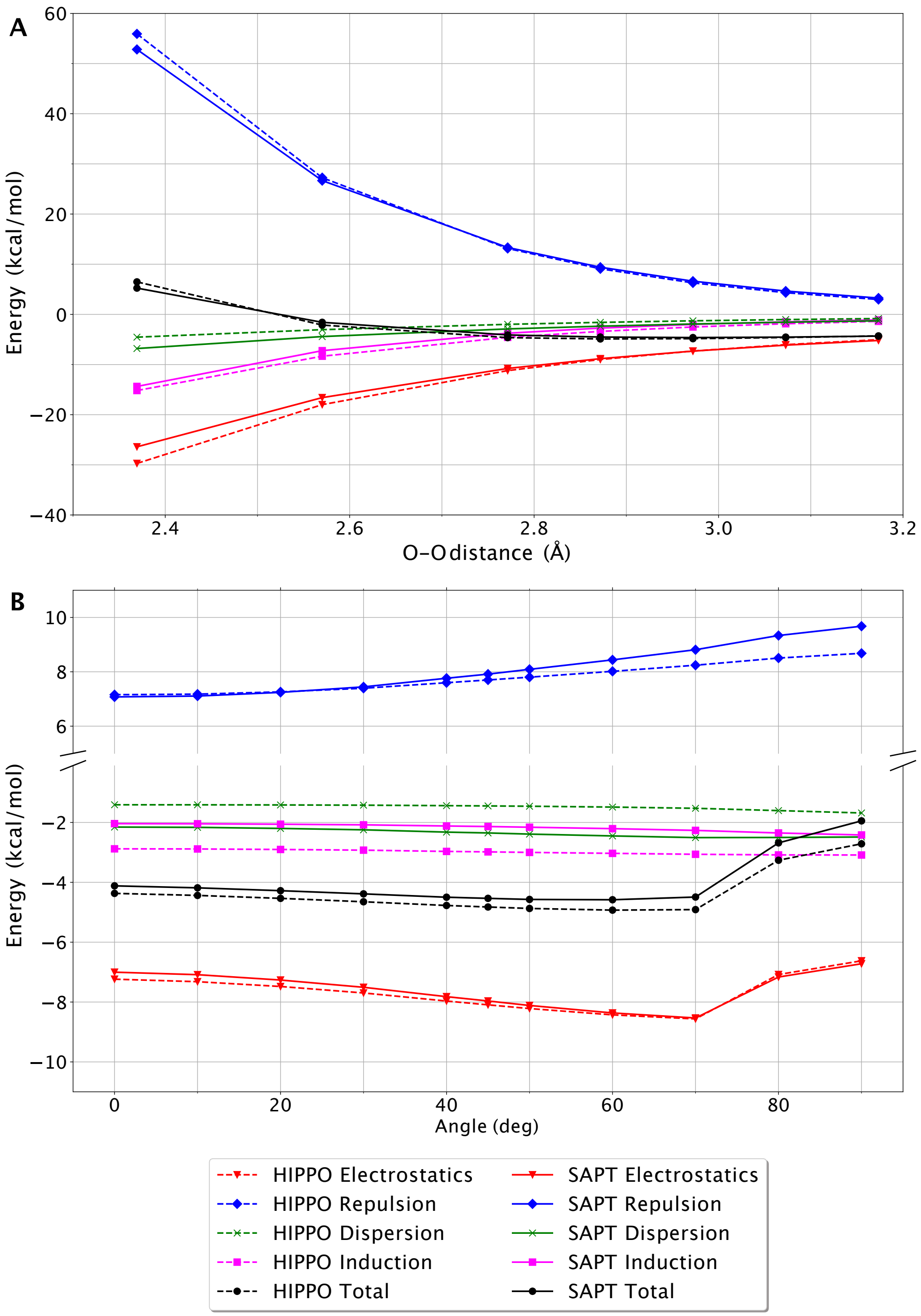

***Figure 3.*** *Energy components for water dimer dissociation curve (A) and "flap angle" degree of freedom (B). HIPPO components are shown in dashed lines and SAPT reference energies are shown in solid. The extent of dissociation is represented by the O-O distance. The "flap angle" is defined as the angle between the O-O vector and the plane of the acceptor water molecule.*

The HIPPO total energy matches the SAPT total energy closely throughout the distance range. Even for the closest points, O-O distances that are rarely sampled in ambient water, the agreement is good. This agreement across the range can be attributed to the fidelity with which HIPPO matches the components of the SAPT energy. In particular, the repulsion and electrostatic curves, which point force fields fail to reproduce at short and long range simultaneously, are in excellent agreement throughout the curve. Importantly, HIPPO is able to capture the short-range physics without compromising the long-range behavior.

Another critical aspect of the water dimer potential energy surface is the hydrogen bond angle. To generate structures for this part of the surface we varied the so-called "flap angle" of the canonical water dimer as illustrated in the inset of Figure 3B. The behavior with respect to this angle is important because it contains information about the anisotropy of the water molecule. Work on the AMOEBA force field has shown that anisotropy in the electrostatics vis-a-vis point multipoles helps reproduce the directionality of hydrogen bonding in water as well as other systems. Here we examine the anisotropy of not just the electrostatics, but the other energy components as well. Plotted in Figure 3B is the change in total energy, as well as the change in each of the components, as the flap angle of the water dimer is changed from $0^o$ to $90^o$. The SAPT curves illustrate an interesting phenomenon. While the dispersion and induction components of the intermolecular energy are largely unchanged across the scan, the electrostatics and repulsion components vary dramatically and in opposite directions. In fact, the trends in these two components counterbalance each other. The minimum of the electrostatic curve lies near 70°. However, the optimal hydrogen angle for the water dimer is known to be slightly smaller, around 60°.

Figure 3B shows that this is nearly entirely due to the countervailing angular dependence of the repulsion curve. It also shows that HIPPO matches the angular dependence of both the electrostatics and repulsion curves well. The anisotropy in the repulsion curve is noteworthy since this is the first force field to include multipolar anisotropic repulsion. This gives a flap angle for the minimized HIPPO water dimer of 63°, near the experimental value of 57°. This underscores the importance of including anisotropy, not just in the electrostatics portion of the force field, but the repulsion as well. Without the angular dependence of the multipolar repulsion model, as is the case in the vast majority of isotropic Lennard-Jones van der Waals functions, the flap angle of the water dimer would be incorrect. Curiously, the original multipole-based AMOEBA model corrected this issue empirically by scaling down the quadrupole moments of each atom by a factor of 0.73, but misdiagnosed the problem. The key to capturing the anisotropy of the potential energy surface of the water dimer seems to be in including anisotropy in the repulsion as well.

The final piece of the dimer potential energy surface we examined is the ten Smith water dimers. These dimers are all stationary points on the water dimer potential energy surface and as such, they form a representative sample of the various dominant dimer configurations in the condensed phase. There are a variety of both hydrogen bonded and non-hydrogen bonded structures in the set, making it a good test of the accuracy of the model with relevant contact geometries beyond the canonical configuration.[66] Fully optimized *ab initio* structures at the MP2/aug-cc-pV5Z level were computed as part of the present study, and are depicted in Figure 4. From the geometry of each dimer at the MP2/aug-cc-pV5Z level, we then determined a “gold standard” counterpoise-corrected CCSD(T) total stabilization energy for each dimer compared to the energy of two optimized, separated monomers at the same level of theory.[67] Note that these energies contain the deformation energy of the water monomers upon dimer formation. The coordinates of the optimized Smith dimers are provided in Supporting Information. Only dimer 1 is a true minimum on the potential energy surface, while the other dimers have one to three negative Hessian eigenvalues.

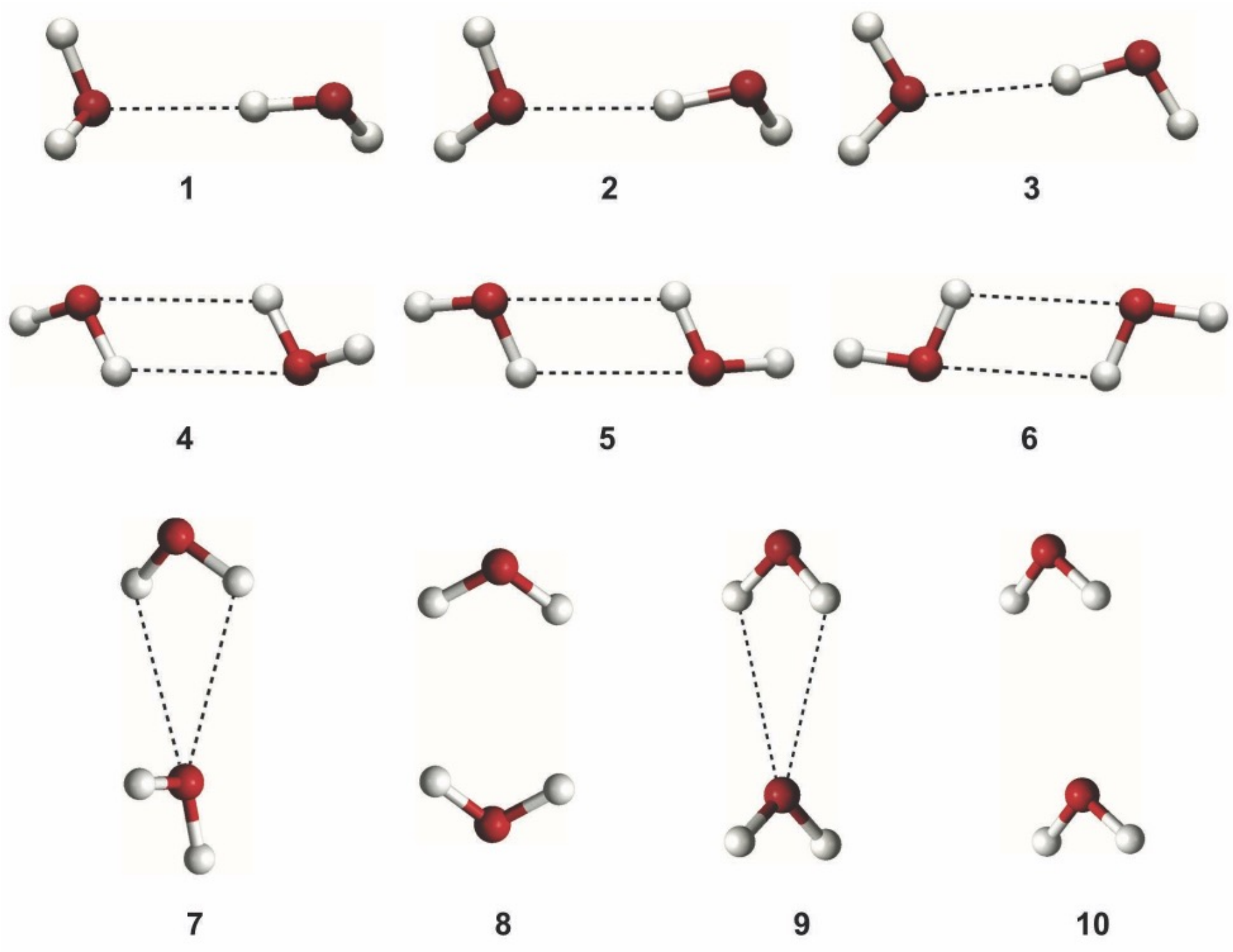

***Figure 4.*** *Structures of the ten Smith water dimers obtained from full geometry optimization at the MP2/aug-cc-pV5Z level. The dashed lines represent hydrogen-oxygen interactions that are roughly within the distance corresponding to the hydrogen bonding. Dimers 1-3 each contain a single hydrogen bond, and are variations of the global minimum structure 1. Dimers 4-6 contain two hydrogen bonds between a pair of antiparallel O-H bonds. Dimer 7, 9 and 10 have two weaker hydrogen bonds of approximately equal distance provided by a single donor water. Dimer 8 has stacked, displaced molecules with H-H interactions as the closest contacts. Atomic coordinates are provided in Supporting Information.*

In Table 2, the structures and energetics of the Smith dimers optimized with the HIPPO model are compared against *ab initio* reference data.[66, 68] In addition to previously reported reference energy values, the MP2/aug-cc-pV5Z structures computed here were used to generate CCSD(T)/CBS energies for all ten dimers. The root mean square energy difference between the CCSD(T)/CBS and HIPPO values is 0.129 kcal/mol, and the average structural RMS with all atoms weighted equally is 0.075 Å. Overall, the structural and energetic agreement is excellent. Dimers 4 and 5 exhibit the largest deviation between QM and HIPPO results. In both cases, the HIPPO optima have lower energies and smaller intermolecular contact distances, perhaps due to a small error in the interaction between antiparallel O–H bonds. The energies of dimers 7 to 10 differ the most between the earlier rigid monomer interaction energies of Tschumper, *et al.,* [66] and the fully flexible values reported by Wang and Bowman[68] or the flexible CCSD(T) values reported here. Unsurprisingly, three of those dimers exhibit the largest deformation energies upon dimer formation. Comparison of HIPPO energies with a limited set of other empirical water models is detailed in Table S2 of the Supporting Information.

***Table 2.*** *Water dimer binding energies for HIPPO compared to ab Initio reference calculations. Dimer geometries were taken from the Supporting Information of reference 69; Ref 1 energies are from reference 66, and Ref 2 values are from reference 68. Dimer stabilization energies[67] and total deformation energies at the CCSD(T)/pV5Z level are shown, as are complete basis set (CBS) extrapolated values.[70, 71] HIPPO dimer energies are provided for single point calculations at the CCSD(T)/pV5Z geometry, and for fully optimized HIPPO structures. Also shown are the QM and HIPPO $R_{O\text{-}O}$ dimer distances, the HIPPO structure RMS vs. MP2/aug-cc-pV5Z optima, and the number of negative frequencies ($\nu$) for CCSD(T)/pV5Z and HIPPO optima. All energies are in kcal/mol, and the $R_{O\text{-}O}$ distance and HIPPO RMS values are in Angstroms.*

| Dimer | Ref 1 | Ref 2 | CCSD(T) /pV5Z | Deform | CCSD(T) /CBS | CCSD(T) $R_{O\text{-}O}$ | HIPPO (sngl) | HIPPO (opt) | #neg ν | HIPPO RMS | HIPPO $R_{O\text{-}O}$ |
|---|---|---|---|---|---|---|---|---|---|---|---|
| 1 | -4.968 | -4.98 | -4.956 | 0.041 | -4.967 | 2.895 | -4.917 | -4.957 | 0 | 0.054 | 2.884 |
| 2 | -4.453 | -4.45 | -4.447 | 0.038 | -4.459 | 2.905 | -4.330 | -4.339 | 1 | 0.104 | 2.913 |
| 3 | -4.418 | -4.38 | -4.398 | 0.037 | -4.410 | 2.911 | -4.232 | -4.238 | 2 | 0.017 | 2.916 |
| 4 | -4.250 | -4.23 | -4.262 | 0.029 | -4.281 | 2.800 | -4.378 | -4.574 | 1 | 0.103 | 2.756 |
| 5 | -3.998 | -3.97 | -4.014 | 0.032 | -4.034 | 2.771 | -3.994 | -4.193 | 1 | 0.161 | 2.754 |
| 6 | -3.957 | -3.91 | -3.969 | 0.036 | -3.991 | 2.748 | -3.823 | -3.913 | 3 | 0.044 | 2.729 |
| 7 | -3.256 | -3.15 | -3.157 | 0.092 | -3.168 | 2.952 | -3.090 | -3.121 | 2 | 0.028 | 2.917 |
| 8 | -1.300 | -1.46 | -1.417 | 0.035 | -1.425 | 3.325 | -1.354 | -1.377 | 3 | 0.046 | 3.271 |
| 9 | -3.047 | -3.18 | -3.197 | 0.114 | -3.208 | 3.018 | -3.169 | -3.184 | 1 | 0.031 | 2.971 |
| 10 | -2.182 | -2.28 | -2.275 | 0.096 | -2.286 | 3.168 | -2.278 | -2.295 | 2 | 0.025 | 3.118 |

Further structural and energetic results, comparing HIPPO with *ab initio* results on the ten dimers, are plotted in Figure 5. The figure shows two levels of comparison. First, it compares the total interaction energy for each dimer. Along with the HIPPO values, two *ab initio* results are shown. The first is the SAPT total energy at the SAPT2+ level. The second *ab initio* values are the CCSD(T)/CBS results obtained in this work. It is interesting to note there is some disagreement between the SAPT and CCSD(T) results. For some dimers, the SAPT value differs by ~0.5 kcal/mol. This shows that although SAPT2+ is useful for determining individual components of the energy function, it is not a replacement for high-level coupled cluster total energy calculations. Optimized HIPPO dimer structures and energies are in good agreement with the CCSD(T) results for all ten dimers. This indicates an accurate balance between the hydrogen bonded and non-hydrogen bonded configurations. The origin of this balance is illustrated by the second level of comparison in Figure 5, the components of the interaction energy. The electrostatics, repulsion, dispersion and induction components of the HIPPO model match the SAPT decomposition in a consistent fashion across the dimer configurations. This demonstrates the agreement in total energies is not coming from cancellation of errors, indicating that a similar agreement should hold for other water dimer configurations outside this set of ten structures.

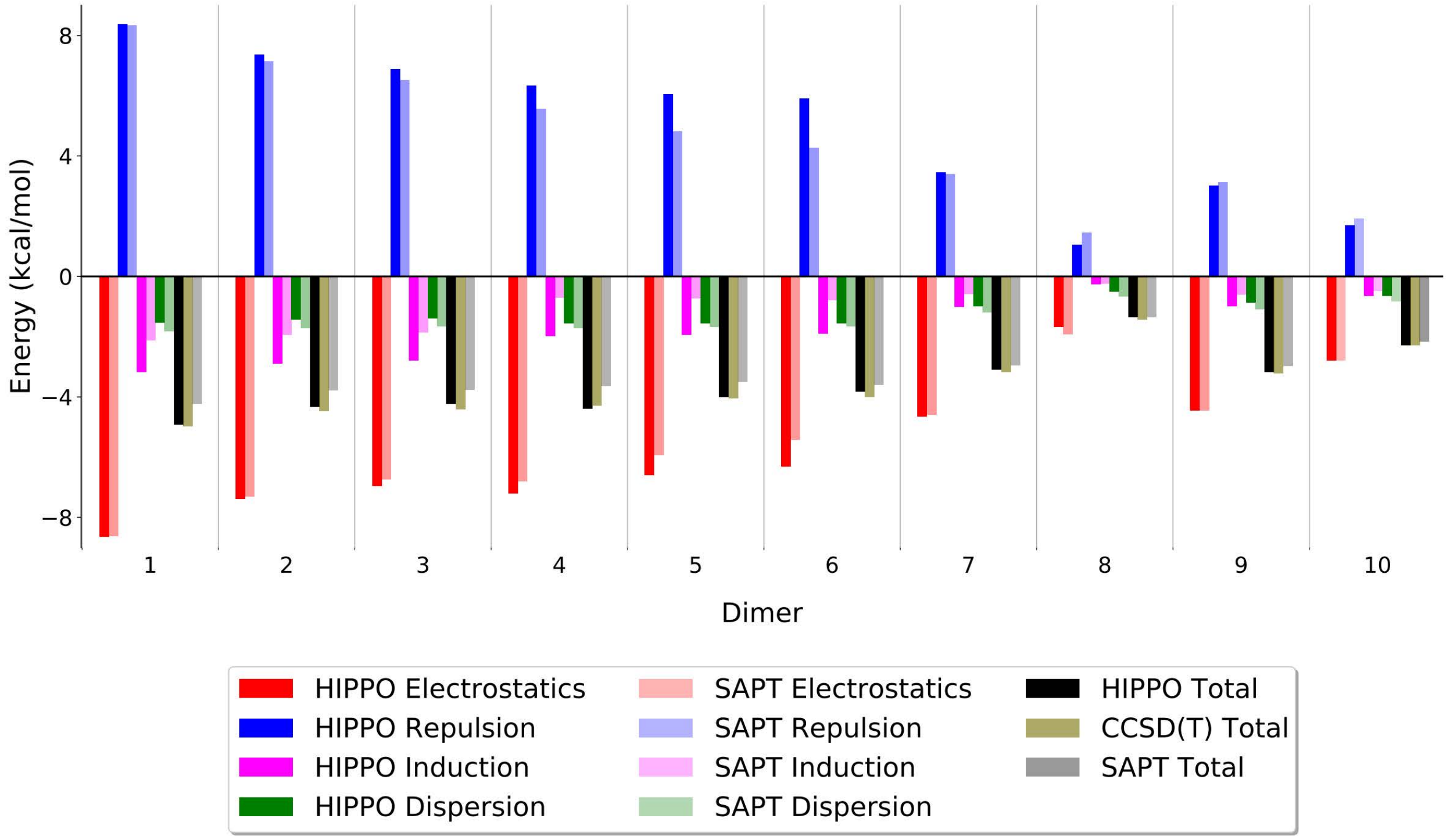

***Figure 5.*** *Total energies and components for ten water dimer stationary points. HIPPO and SAPT components are shown as colored bars. Although all errors are under 0.5 kcal/mol, these show some compensation on the part of HIPPO between the induction and dispersion components. The black, tan, and grey bars represent the HIPPO, CCSD(T), and SAPT values, respectively. Notably, HIPPO is in better agreement with the CCSD(T) data than SAPT, suggesting that the HIPPO component errors relative to SAPT are within the intrinsic error of the SAPT methodology.*

***Larger Clusters.*** We next tested the HIPPO model on larger clusters of water ranging from three to twenty molecules. The goal here is to span as much of the gap as possible between gas phase and condensed phase. For these larger clusters, SAPT data becomes difficult to interpret, but there are two types of data relevant to evaluating the HIPPO model. First, we compare total cluster binding energies. This provides a measure of how well the potential energy function performs as water becomes more liquid-like. Second, we compare the many-body energies. The average dipole moment of a water molecule increases steadily upon moving from monomer to dimer to clusters to condensed phase. This implies that in order for any model to achieve agreement with QM data for both clusters and condensed phase, it must include many-body effects. Thus, we compare the many-body energies from *ab initio* calculations with the many-body energies from the classical HIPPO polarization function.

HIPPO compares very well with gold standard CCSD(T) benchmark total energy calculations moving from gas phase dimers toward bulk-like clusters. As shown in Tables 2 and 3, the agreement for structures through the hexamer is within 0.57 kcal/mol on average. Moreover, the relative ranking of unique structures is also quite accurate. For example, HIPPO ranks the eight reference water hexamer structures in the same order as CCSD(T) calculations. Lastly, the HIPPO minima are structurally very similar to the reference QM-optimized structures, indicating the accuracy of the local potential energy landscape.

Unlike pairwise force fields, where the total energy of a system is simply the sum of the energies of every pair of interactions, HIPPO is polarizable. This means that it is designed to reproduce the non-additive portion of intermolecular interactions. To quantify the amount of non-additivity, we compute the three-body energy of a range of different water clusters. The three-body energy is defined as

$$E_{3B} = E_{total} - \sum_i \sum_j E_{ij} + \sum_i E_i \tag{25}$$

where second and third terms represent the sums of the two-body and one-body energies, respectively. The four-body and higher terms are negligible in the case of water.[72] The first test set for three-body energies is the water trimer at a range of intermolecular contact distances. Starting from the structure depicted in the inset of Figure 6, the distances $d_1$ and $d_2$ were varied systematically. The three-body energy was computed at the MP2 level of theory and compared to the HIPPO three-body energy. Figure 6 shows that across the range of distances HIPPO agrees well with the *ab initio* result. Particularly at distances near equilibrium the agreement is very good.

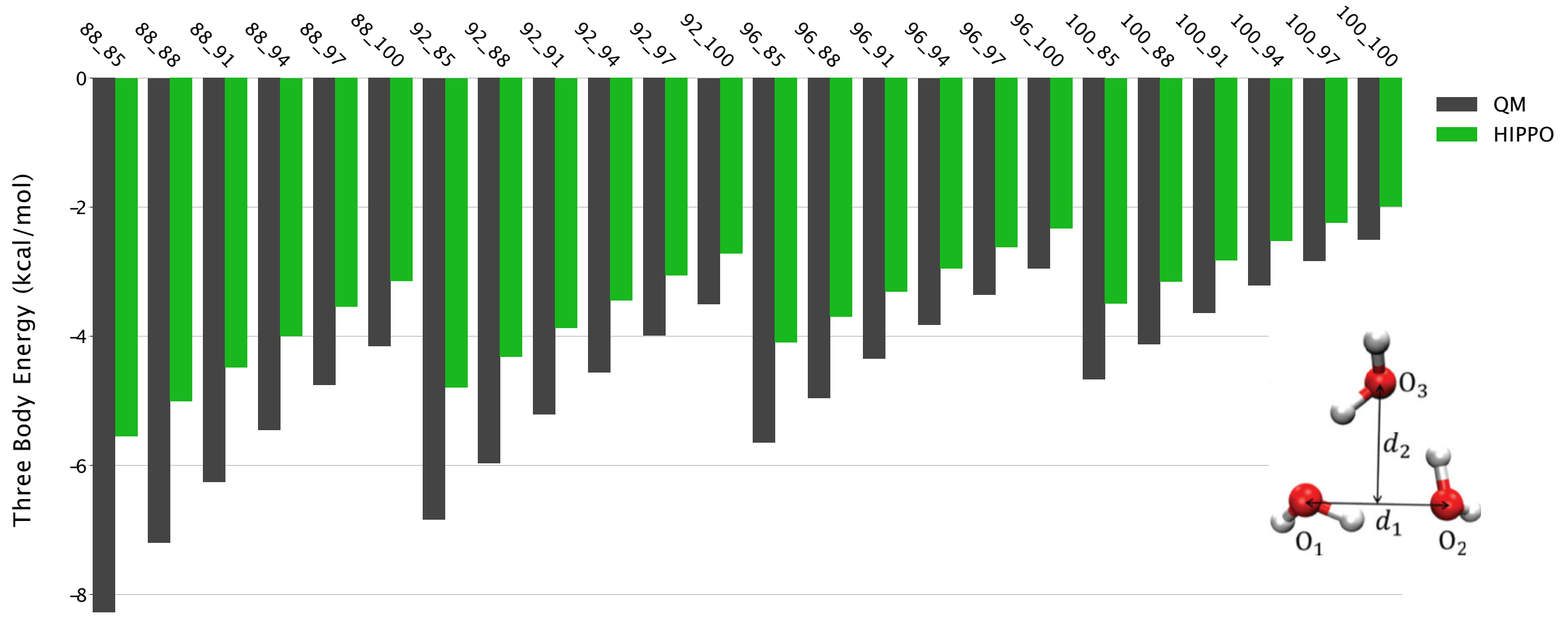

***Figure 6.*** *Three-body energies for water trimer as a function of intermolecular distance. HIPPO is within 1 kcal/mol for near-equilibrium structures. The X-axis values represent $d_1$ and $d_2$, respectively as shown in the inset as percentages of the equilibrium distances. QM data is generated at the MP2/aug-cc-pVTZ level.*

***Table 3.*** *Water cluster binding energies (kcal/mol) with HIPPO compared to ab initio calculations. [a] Structures and reference values from reference 73. [b] Hexamer structures and reference energies from reference 74.*

| Cluster | Structure | Reference | HIPPO | HIPPO Minimum | Difference |
|---|---|---|---|---|---|
| Trimer [a] | | -15.77 | -15.417 | -15.767 | 0.00 |
| Tetramer [a] | | -27.39 | -25.695 | -26.685 | 0.71 |
| Pentamer [a] | | -35.90 | -32.994 | -34.582 | 1.32 |
| Hexamer [b] | Prism | -45.92 | -44.169 | -46.145 | -0.23 |
| | Cage | -45.67 | -43.635 | -45.387 | 0.28 |
| | Bag | -44.30 | -41.106 | -43.364 | 0.94 |
| | Chair | -44.12 | -40.484 | -42.543 | 1.58 |
| | Book A | -45.20 | -42.359 | -44.245 | 0.96 |
| | Book B | -44.90 | -42.103 | -43.958 | 0.94 |
| | Boat A | -43.13 | -39.576 | -41.548 | 1.58 |
| | Boat B | -43.07 | -39.612 | -41.555 | 1.52 |
| Octamer [a] | $D_{2d}$ | -73.0 | -68.309 | -71.547 | 1.5 |
| | $S_4$ | -72.9 | -68.253 | -71.559 | 1.3 |
| 11-mer [a] | 43’4 | -104.6 | -94.775 | -100.232 | 4.4 |
| | 515a | -1040 | -93.635 | -99.377 | 4.6 |
| 16-mer [a] | AABB | 164.1 | -155.457 | -161.556 | 2.5 |
| | ABAB | 164.2 | -155.875 | -161.836 | 2.4 |
| | Antiboat | 164.6 | -152.799 | -159.634 | 5.0 |
| | Boat A | 164.4 | -152.457 | -159.357 | 5.0 |
| | Boat B | 164.2 | -152.400 | -159.425 | 4.8 |
| 17-mer [a] | 552’5 | -175.7 | -161.740 | -169.938 | 5.8 |
| | Sphere | -175.0 | -162.549 | -170.681 | 4.3 |
| | | | | | |
| | | | | **MAD:** | |
| **Summary:** | Dimer – Hexamer | | | 0.57 | |
| | Octamer – 17-mer | | | 3.8 | |

This level of agreement illustrates two important points. First, it shows that the HIPPO model is effective in capturing the many-body effect, and thus may perform well across the spectrum from gas to condensed phase. Second, it suggests that the majority of the *ab initio* many-body energy can be classified as polarization. It is well known that other categories of intermolecular interaction such as dispersion, charge transfer and repulsion have many-body components. The data in Figure 6 shows that for water, however, these appear to be small. The HIPPO three-body energy is systematically smaller than the *ab initio* result, but only by a small amount. It is only at the closest points, where water rarely accesses in the condensed phase, that it appears that higher-order many-body effects start to be significant.

To assess if the agreement with small-scale *ab initio* many-body results translates to liquid water, we also computed the three-body energy of progressively larger clusters. For water clusters of four to eight molecules, we computed the three-body energy at the MP2 level of theory and compared it against HIPPO results. Figure 7 shows the trends seen in the trimer test case hold for larger clusters.

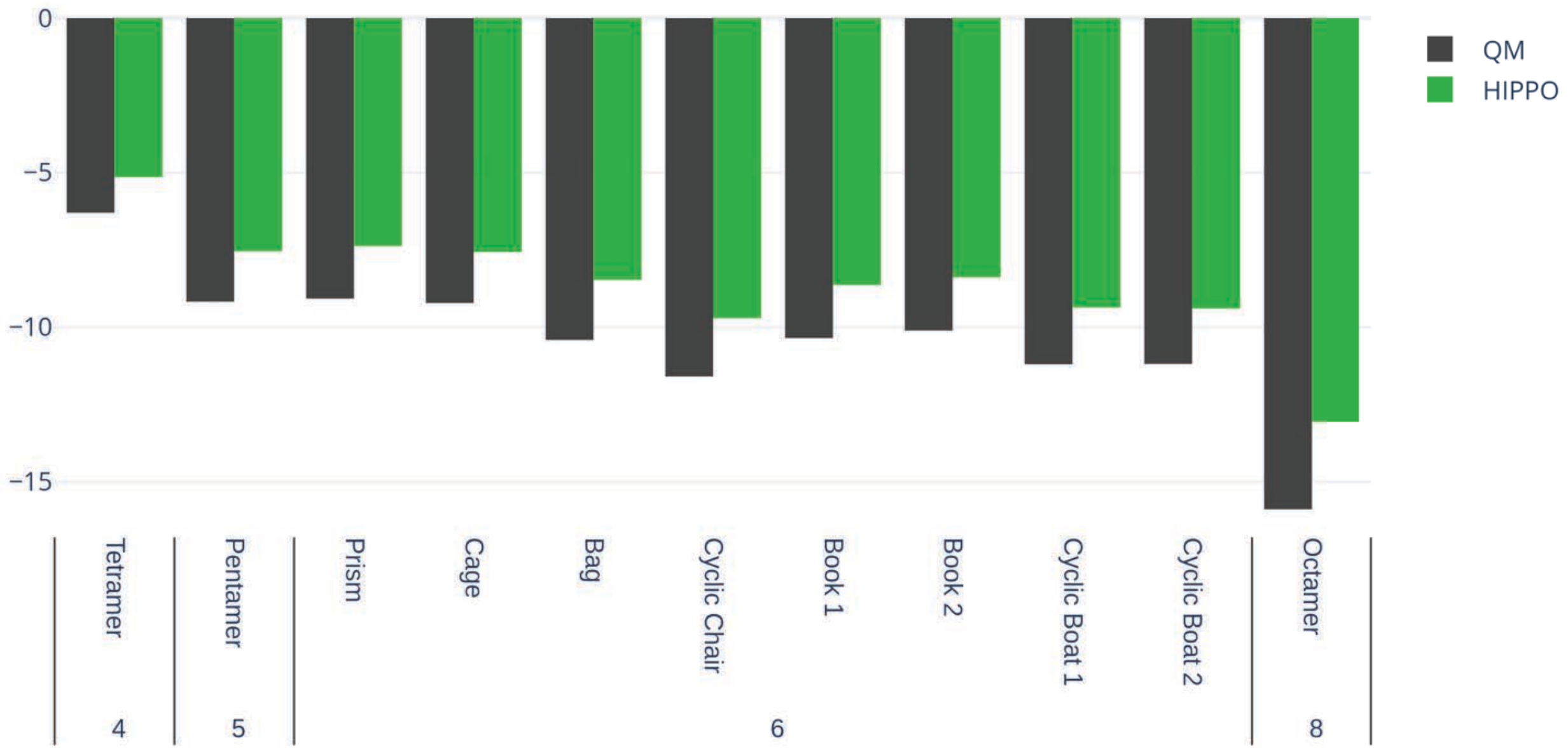

***Figure 7.*** *Three-body energies for water clusters tetramer through octamer. QM data is generated at the MP2 level of theory and aug-cc-pVTZ basis set.*

Just as in the trimer case, the HIPPO result always slightly underestimates the magnitude of the total *ab initio* three-body energy. This validates the observation from the trimers that many-body effects in higher-order terms appear to stay small as cluster size grows. Additionally, the behavior of the three-body energy with geometry appears to be in good agreement with the reference data. HIPPO correctly predicts the ordering of the amount of three-body energy in the eight water hexamer structures. These structures are picked to be representative of fully hydrated water. It is difficult to estimate what the many-body energy of a full, condensed phase system of water is, but this result for the hexamers suggests that HIPPO may give an adequate representation.

***Liquid Properties.*** In addition to accurately modeling *ab initio* data, it is important for a water model to accurately reproduce experimental liquid phase properties as well. We have tested the HIPPO water model on a wide variety of experimental observables at room temperature and ambient pressure and present the results in this section.

***Table 4.*** *Water properties at room temperature (298 K and 1 atm). [a] Reference 75. [b] Reference 76. [c] Reference 77. [d] Reference 78. [e] Reference 79. [f] Reference 80.*

| Property | HIPPO | Experimental | Abs. Deviation |
|---|---|---|---|
| Density ($kg/m^3$) | 996.492 | 997.045[a] | 0.553 (0.06%) |
| Enthalpy of Vaporization (kJ/mol) | 43.806 | 43.989[b] | 0.183 (0.42%) |
| Static Dielectric Constant | 76.878 | 78.409[c] | 1.531 (1.95%) |
| Self-Diffusion Coefficient ($10^{-5}$ $cm^2/s$) | 2.557 | 2.299[d] | 0.258 (11.22%) |
| Surface Tension ($mJ/m^2$) | 74.918 | 71.99[e] | 2.928 (4.07%) |
| Second Virial Coefficient (L/mol) | -1.2612 | -1.158[f] | 0.103 (8.91%) |

The primary thermodynamic and dynamic properties of the HIPPO model are collected in Table 4 along with the known experimental values. The density is in excellent agreement with experiment, with an error of less than 0.1%. The heat of vaporization, or the amount of energy required to transfer a water molecule from the liquid phase to the gas phase, is also in excellent agreement with an error relative to experiment of 0.4%. Both of these values were included in the objective function of the parameter refinement step, so good agreement is expected.

Likewise, the dynamic properties of HIPPO water are in close agreement with experiment. The self-diffusion coefficient of water measures how quickly or freely water molecules move in the liquid phase. The predicted diffusion coefficient of HIPPO differs from the experimental value by 11%. This is a reasonable agreement for a quantity that is known to be quite sensitive to details of molecular dynamics simulations. The HIPPO model is also in excellent agreement with the experimental dielectric constant of water. This is also a highly sensitive quantity for molecular dynamics simulations, and the HIPPO prediction is within 2% of the experimental result. The agreement with the experimental dielectric constant indicates that the HIPPO water electrostatic environment is accurate. This is important not just for the properties of water, but for the future use of the HIPPO model solvating small molecules, ions and ultimately biological macromolecules. Lastly, the surface tension, a stress test for how well a water model handles the balance between bulk solution and interfaces, of the HIPPO model is in excellent agreement with experiment. The accuracy in the surface tension suggests that HIPPO will model solvation of both polar and hydrophobic species equally well.

The structural properties of liquid water are also of great interest for both the study of pure water and water as a solvent. As the canonical example of the hydrogen bond and because of its bent shape, liquid water represents a balance between many orientations of water-water interactions. To probe these structural properties, we compared the experimental radial distribution functions and second virial coefficient of water to those predicted by HIPPO.

Plotted in Figure 8 are the O-O, O-H, H-H radial distribution functions of water. Panel A in Figure 8 shows that the O-O radial distribution function of HIPPO water is in good agreement with experiment. The position of the first peak at 2.785 Å and height at 3.0 is within the experimental uncertainty of the experimental curve. The entire curve lies within the "family" of O-O g(r) curves described by Brookes and Head-Gordon.[81] The O-O g(r) represents the coarse molecular level of structure in liquid water. The close agreement of HIPPO shows that the force field has the correct number of molecules in each solvation shell. At a finer level of detail, the O-H and H-H curves are also in close agreement with experiment. Panels B and C of Figure 8 show that the positions of the peaks in these curves agree with experiment. Moreover, the relative heights of the first and second peaks in the O-H and H-H curves correspond closely to the relative heights from the experimental model. This suggests that not only are the correct number of molecules in each solvation shell, but the average orientations of those molecules are in line with reality as well.

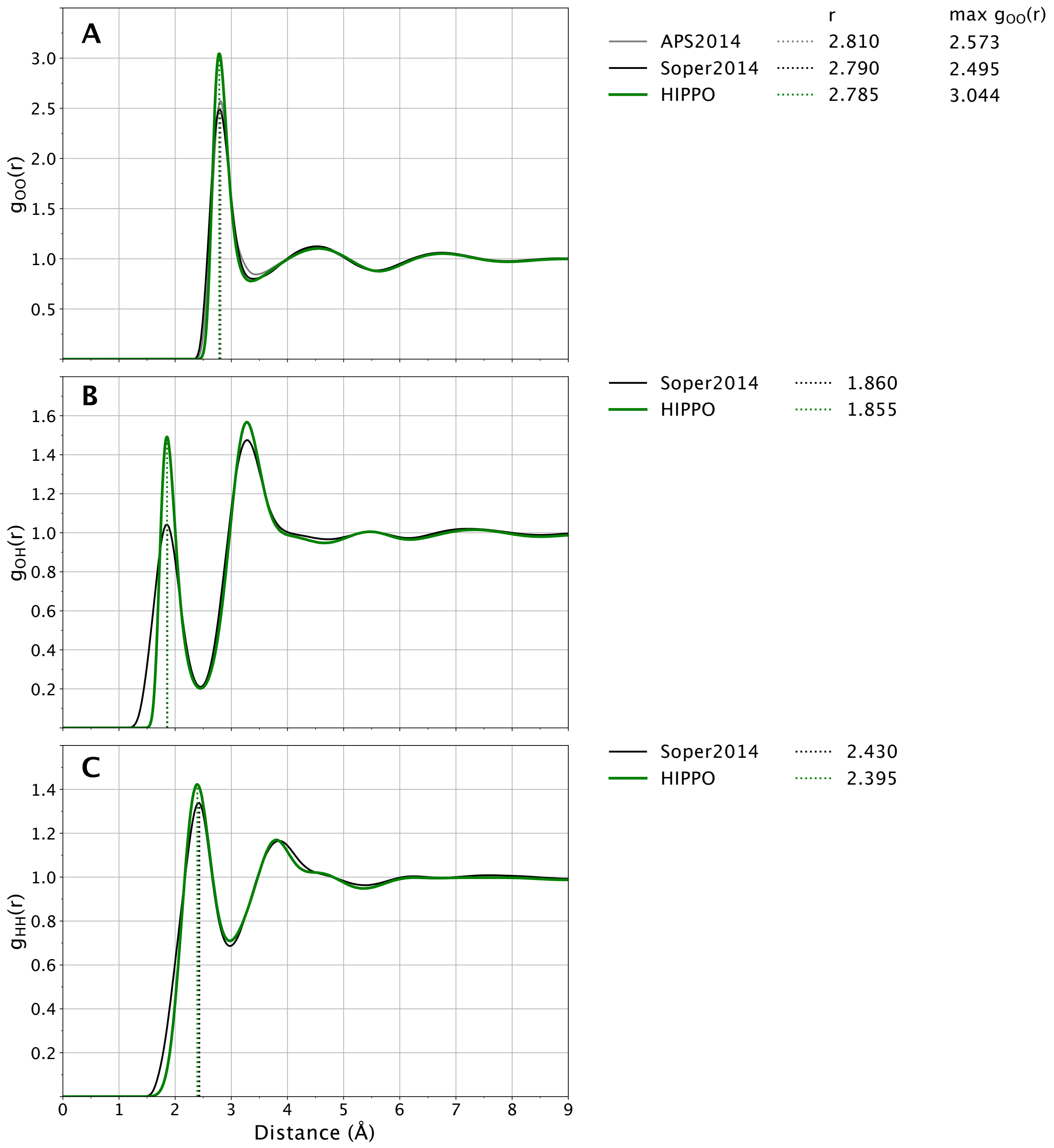

***Figure 8.*** *Water radial distribution at 298 K and 1 atm. HIPPO results are shown in green and experimental in black. First peaks of the HIPPO distribution are indicated with dotted green vertical lines. Experimental curves from references 81-83.*

***Temperature and Pressure Dependence.*** To stress the water model, we also tested the HIPPO model at a range of temperatures and pressures. This data is included to evaluate how well the HIPPO model performs in a variety of conditions away from room temperature and ambient pressure. We calculated the density, enthalpy of vaporization, heat capacity, dielectric constant, self-diffusion coefficient, thermal expansion coefficient, and isothermal compressibility at temperatures ranging from super-cool up to the boiling point. The same was done for the second virial coefficient. We also calculated density at a range of pressures, up to 10,000 atm. Of these, only the density, enthalpy of vaporization and dielectric constant were included in the fitting procedure. The results are presented in Figure 9 to Figure 14, and are discussed in detail below.

The temperature dependence of the density of liquid water is unique. At high temperatures, the dependence is intuitive and straightforward: the higher the temperature, the lower the density. At lower temperatures, near the freezing point, however, the correlation is less intuitive. The curve "turns over" and the density starts to decrease as the temperature decreases. This dual dependence leads to the characteristic "temperature of maximum density" of water. Panel A of Figure 9 shows that HIPPO reproduces the entirety of this curve with exceptional accuracy. The error relative to experiment is less than 1% for all points on the curve. The HIPPO temperature of maximum density is 277 ± 2 K, which is in near-perfect agreement with the experimental value of 277 K.

The enthalpy of vaporization temperature dependence is simple. As temperature increases, the $\Delta H_{vap}$ decreases. This matches our intuitive understanding of how much energy it takes to remove a water model from the liquid phase. Panel B of Figure 9 shows that HIPPO exhibits this same behavior, but the slope of the curve is slightly steeper than experiment. The result is a near-perfect enthalpy of vaporization at room temperature with errors of ± 3% at the respective ends of the tested temperature spectrum.

The heat capacity of the HIPPO water model is plotted in panel D of Figure 9. Heat capacity is closely related to the derivative with respect to temperature of the enthalpy of vaporization. Since the slope of the enthalpy of vaporization shown in panel B is largely unchanged over the temperature range, the heat capacity is nearly a constant with respect to temperature. However, since the slope of the HIPPO model for the enthalpy of vaporization in panel B is too steep, the calculated heat of vaporization is noticeably higher than experiment. This difference is the result of a known shortcoming in all classical models of water: the neglect of nuclear quantum effects (NQEs). Rough corrections of 6 cal/mol/K and 2 cal/mol/K have been suggested.[84, 85] The ForceBalance program also implements an NQE correction for the enthalpy of vaporization,[86, 87] and these corrected values are plotted in panel B of Figure 9. Analysis of the NQE correction and its ramifications are discussed in greater detail in the Discussion section below.

As a model whose intended future use is the solvation of biological macromolecules, the dielectric constant is of great importance. One of the main practical implications of using a polarizable water model in biomolecular simulations is accurately modeling both water in bulk solvent and isolated water molecules in, for instance, a protein binding pocket. We calculated the dielectric constant of the HIPPO model and the results are plotted in panel A of Figure 10. The dielectric constant is notoriously sensitive and difficult to converge. However, the HIPPO model shows good agreement across the range of temperatures. This stands in contrast to most fixed charge water models whose dielectric constants change very little with temperature.[88]

Another typically sensitive property of water models is the self-diffusion coefficient. This property is also important to future biomolecular simulations because it is a contributing factor to producing accurate timescales for macromolecular dynamics. Plotted in panel B of Figure 10 is the self-diffusion coefficient of HIPPO water *vs*. temperature. It is clear that the overall shape of the temperature dependence curve is correct, with the HIPPO diffusion slightly higher than experiment. The self-diffusion coefficient is a rough measure of the balance of hydrogen bonding *vs.* other types of intermolecular interactions in water. The agreement of HIPPO with experiment indicates this balance is accurate. Due to the steep rise in diffusion coefficient with temperature, the 11% overestimation by HIPPO at room temperature corresponds to only a small error along the temperature dimension. For example, the computed HIPPO coefficient of 2.557 ± 0.026 $\times 10^{-5}$ $cm^2/s$ at 298 K is equivalent to the experimental value at roughly 304 K. Figure 11 shows the dependence of the diffusion coefficient on the reciprocal dimension of the cubic simulation box, 1/L. The variation with box size is in agreement with the well-known correction suggested by Yeh and Hummer.[63] Since the Yeh-Hummer correction depends on the shear viscosity of each model, we feel a diffusion *vs.* 1/L plot provides the best diffusion estimate at infinite box size for any specific water model. The HIPPO value obtained from Figure 11 is $2.568 \times 10^{-5}$ $cm^2/s$ at 298 K, which is very close to the average of $2.557 \times 10^{-5}$ $cm^2/s$ from multiple 100 Å cubic box simulations.

Finally, we show how the HIPPO water model performs under extreme conditions. Panels C and E show the temperature dependence of the thermal expansion coefficient and isothermal compressibility, respectively. The HIPPO compressibility is higher than the experimental value. This agrees with the pressure dependence of the density shown in Figure 13, where the density is greater than predicted for high pressure simulations. Note, however, the units of compressibility are small. Water is very difficult to compress and the HIPPO model of water is only slightly less so. The agreement of the thermal expansion coefficient with experiment is better. Cold water expands rapidly as it is heated up, but the rate of expansion slows as the temperature increases. HIPPO reproduces this trend, mirroring the behavior seen in the density *vs*. temperature curve.

***Ice Properties.*** In addition to liquid properties, we tested the properties of ice crystals. Due to its variety of structures, ice is a stringent test of the intermolecular potential. The intermolecular distances are generally shorter than in liquid water and thus stress the repulsive wall of the model. We computed lattice energies and densities for ten different ice polymorphs across a range of conditions. The HIPPO results are shown against curated experimental data in Table 5. Predicted densities are in error by no more than 2.5% and the lattice energies are all within 3% of the experimental values.

***Table 5.*** *Ice properties from HIPPO model compared with experimental density (kg/m$^3$) and lattice energy (kcal/mol).* [a] *Experimental values from reference 89.* [b] *Values from the ICE10 data set.*[90]

| | **Density** | | | | **Lattice Energy** | | | |
|---|---|---|---|---|---|---|---|---|
| | **HIPPO** | **Expt** [a] | **Diff** | **% Error** | **HIPPO** | **Ref** [b] | **Diff** | **% Error** |
| Ice XI | 949.9 | 934 | 15.9 | 1.7 | -13.804 | -14.10 | 0.3 | -2.1 |
| Ice Ih | 910.8 | 920 | -9.2 | -1.0 | -13.699 | -14.07 | 0.4 | -2.8 |
| Ice IX | 1164.8 | 1194 | -29.2 | -2.4 | -13.769 | -13.97 | 0.2 | -1.4 |
| Ice XIV | 1316.7 | 1294 | 22.7 | 1.8 | -13.360 | -13.74 | 0.4 | -2.9 |
| Ice XV | 1355.6 | 1364 | -8.4 | -0.6 | -13.365 | -13.48 | 0.1 | -0.8 |
| Ice Ic | 951.8 | 931 | 20.8 | 2.2 | | | | |
| Ice Ica | 947.2 | 931 | 17.2 | 1.8 | | | | |
| Ice XIII | 1279.1 | 1251 | 28.1 | 2.2 | | | | |
| | | **MAD:** | 18.9 | | | **MAD:** | 0.3 | |

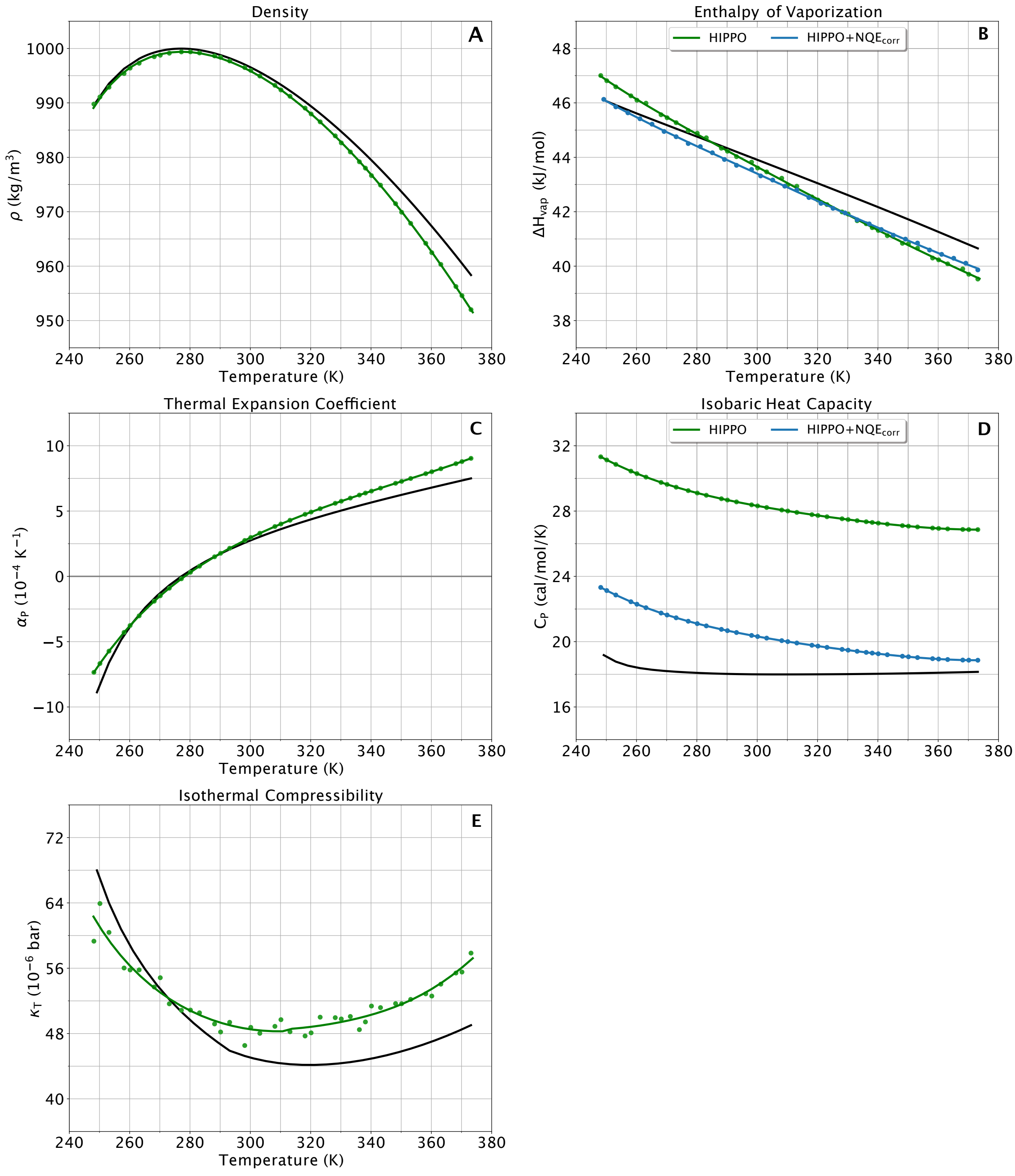

***Figure 9.*** *Thermodynamic water properties for the HIPPO model (green) compared to experiment (black) for temperatures from 248 to 373 K at atmospheric pressure (1 atm). (A), (C), (D), (E) experimental values from reference 75. (B) experimental data from reference 76.*

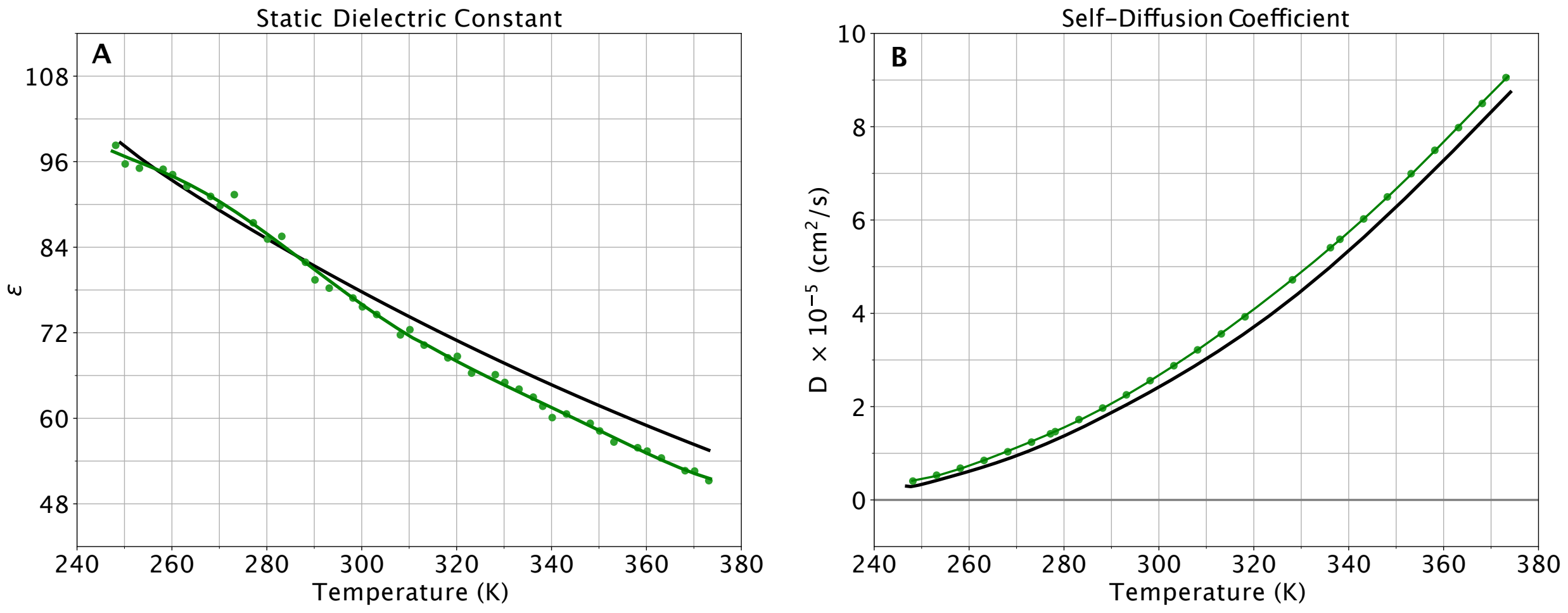

***Figure 10.*** *Dynamical water properties for the HIPPO model (green) compared to experiment (black) for temperatures from 248 to 373 K at atmospheric pressure (1 atm). (A) experimental values from reference 77. (B) experimental values taken from references 78, 91 and 92.*

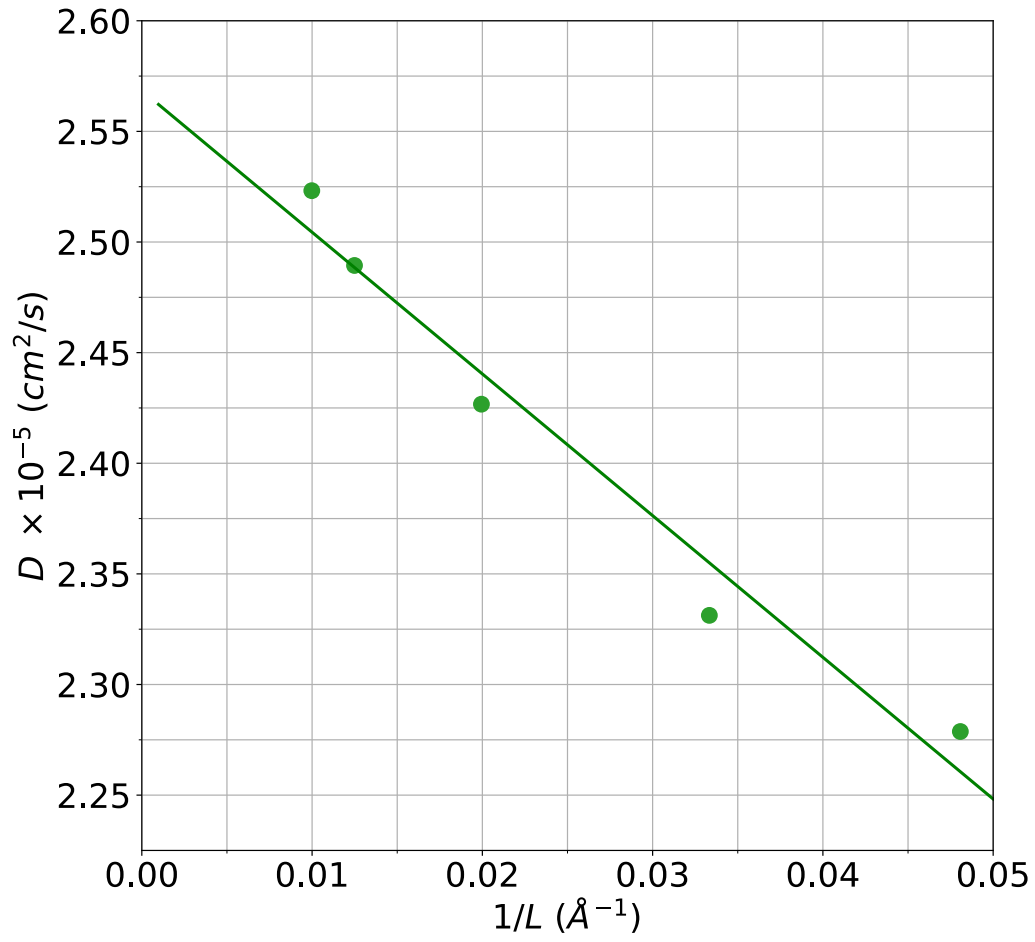

***Figure 11.*** *Self-diffusion coefficient vs. cubic box size at 298 K. The extrapolated y-axis intercept, corresponding to the estimated diffusion coefficient at infinite box size, is* $2.568 \times 10^{-5}$ *cm*$^2$*/s.*

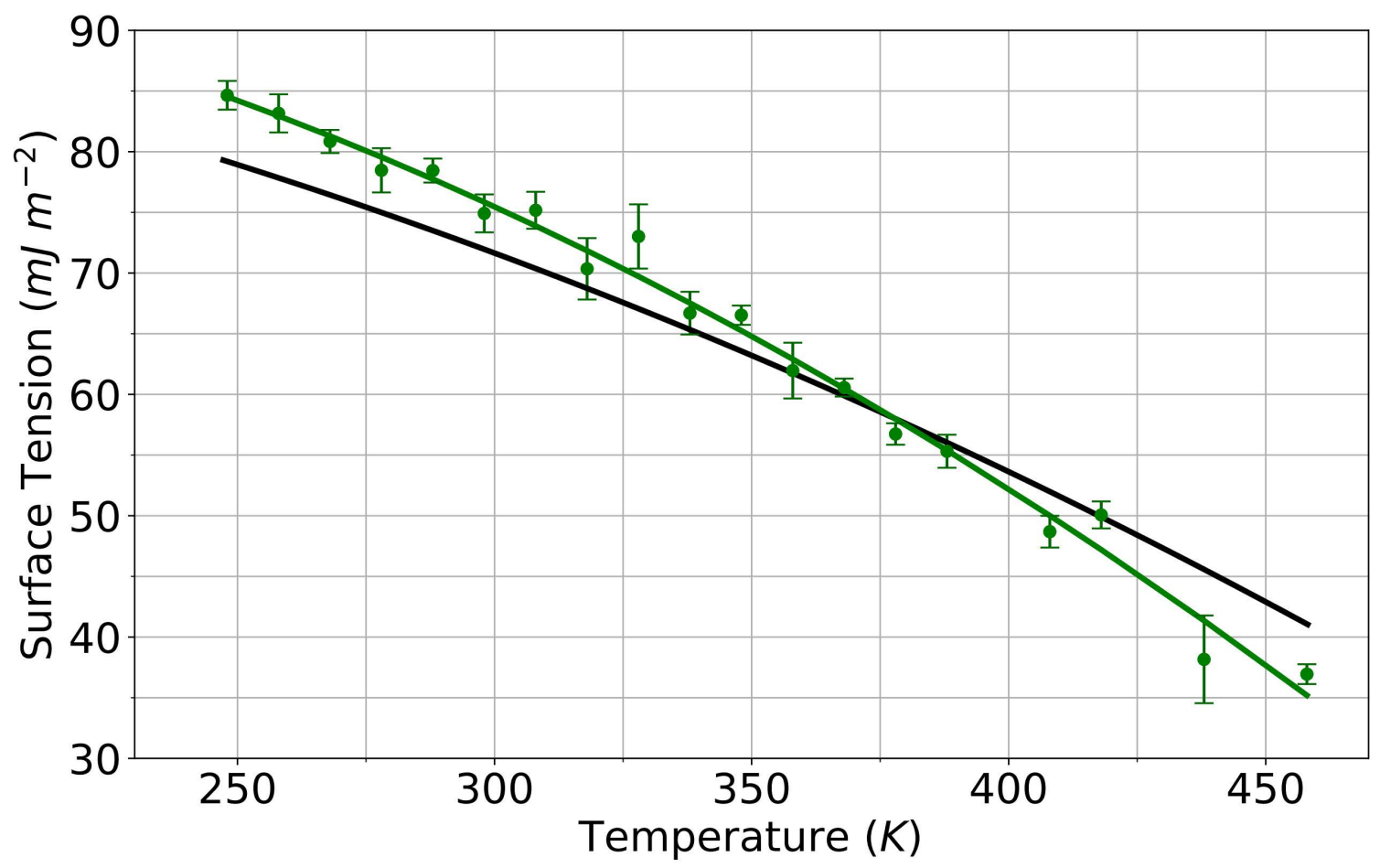

***Figure 12.*** *Surface Tension for the HIPPO model (green) compared to experiment (black) for temperatures from 248 K to 458 K. Experimental values from references 79 and 93.*

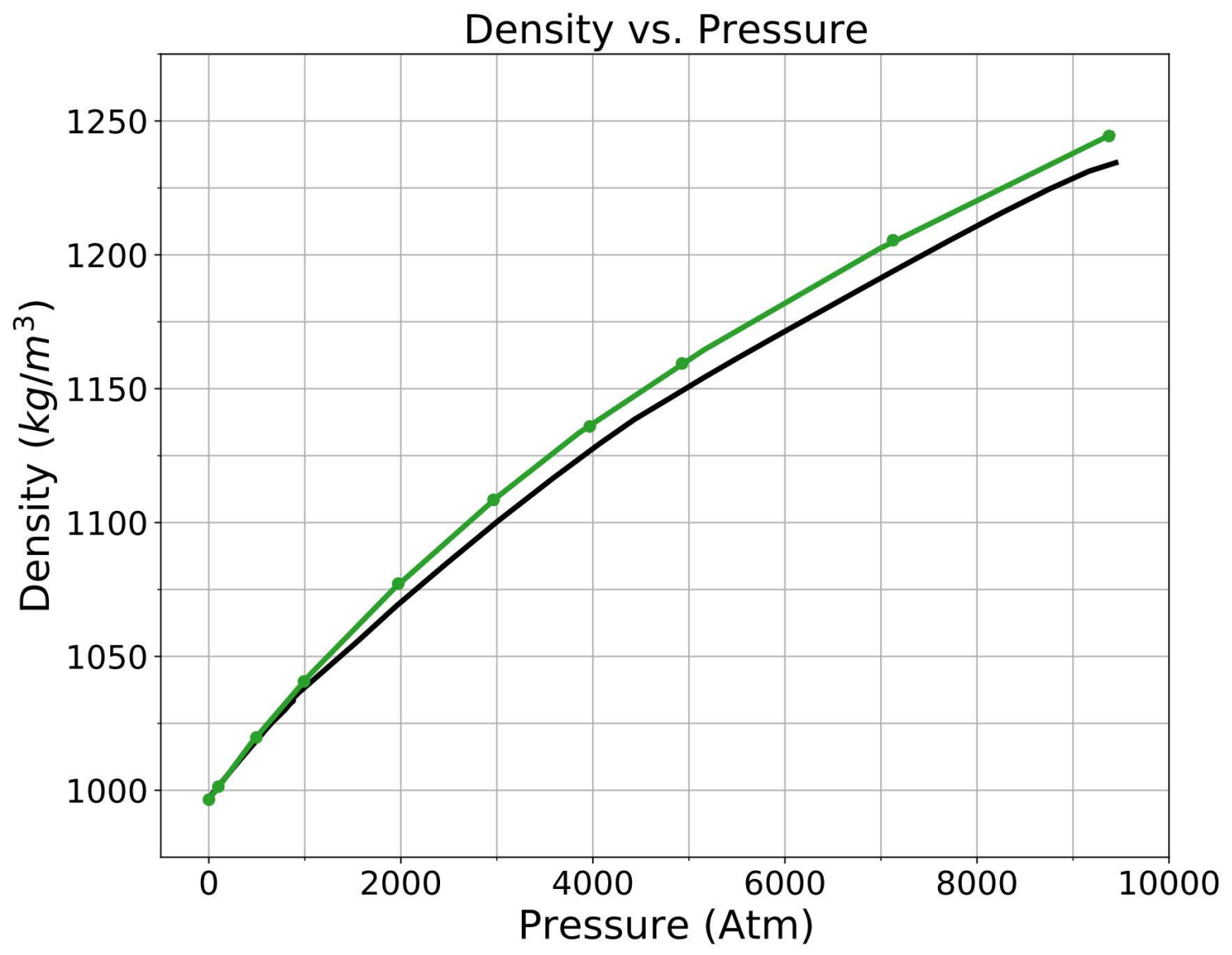

***Figure 13.*** *Density of the HIPPO model (green) compared to experiment (black) for pressures from 1 to 9,375 atm at room temperature (298 K). Experimental values from reference 94.*

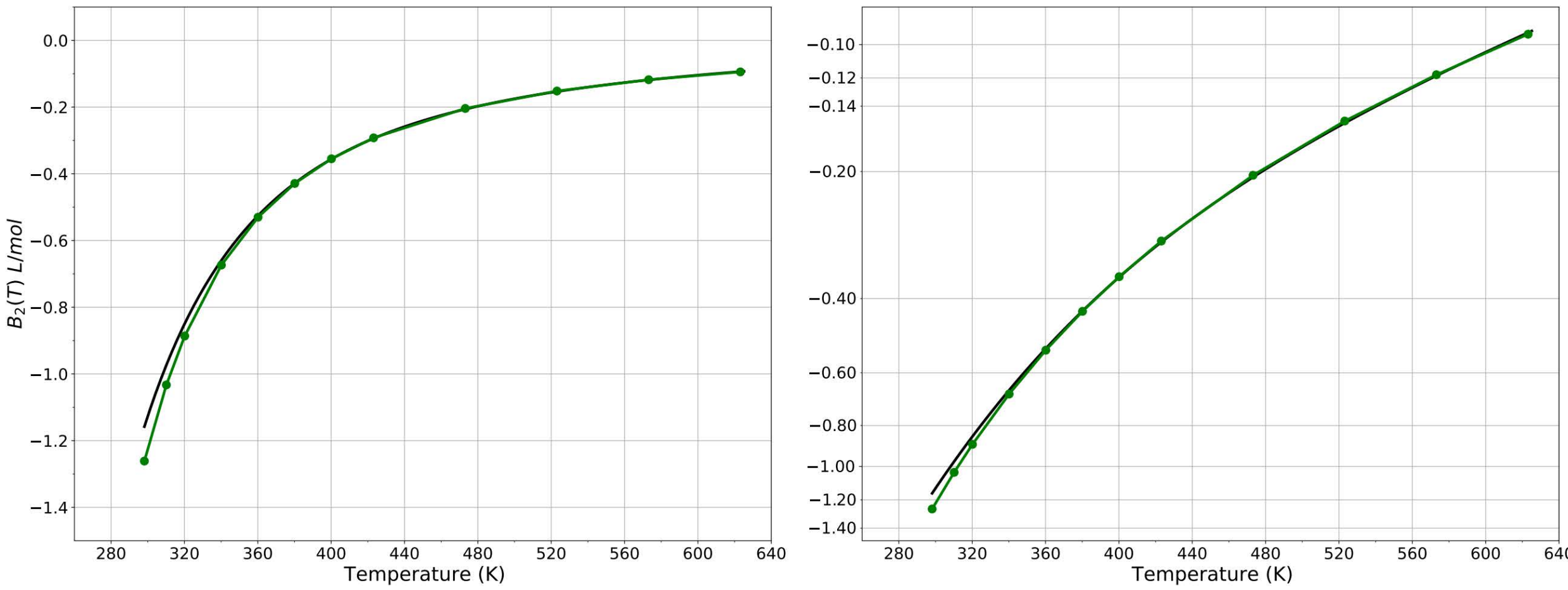

***Figure 14.*** *Second virial coefficient of the HIPPO water model (green) compared to experiment (black) for temperatures from 298 to 575 K. Experimental values from references 80 and 95.*

## Discussion

***Implications of Parameter Space.*** HIPPO is a force field derived from our understanding of how atoms and molecules interact in short range. For this reason, our first goal in building a water model was to guarantee it could accurately reproduce high-level QM calculations for different configurations of water dimers. This is the reason we fit the initial water parameters using the SAPT energy decomposition as a reference for each energy component.

The second and most relevant goal of HIPPO water model is to be appropriate to a variety of MD applications, including solvation of biomolecules. Therefore, the model needs to agree with experimental data available within a small tolerance. The initial parameter set obtained after SAPT fitting did not meet our requirements for the condensed phase properties and led us to continue improving the model.

We chose to perform a constrained global search in parameter space using the repulsion, charge transfer and dispersion parameters of oxygen and hydrogen, centered at their initial values. With the exception of the charge transfer parameters, each parameter was not allowed to vary by any more than 5%. Besides providing improvement of liquid properties, we chose this method because it could give us insight into how the features of the potential energy surface of water dimers related to condensed phase properties with respect to parameter space. Our optimizer was set to only compute liquid properties for parameters that kept the energy of water dimers within an average deviation 0.5 kcal per energy component, compared to the SAPT reference. The flexibility of 0.5 kcal in SAPT component is explained by the fact that SAPT calculations have intrinsic errors compared to gold standard CCSD(T) values. With that requirement, we were able to generate hundreds of water models. This showed we were dealing with a rough potential energy surface and the initial SAPT fitting put the model in a shallow minimum well. Upon computing water properties at room temperature for all the models generated, we selected the one with the smallest combined deviation from condensed phase experimental data and SAPT energy components.

Using the large amount of simulation data generated during parameter optimization, correlation analysis was performed between the energy components of low energy water dimer structures and liquid properties computed for the same set of parameters. Beyond a few obvious exceptions involving repulsion, no clear correlation was seen between calculated liquid properties and dimer total energy or components. Although there is some selection bias in the data, the 0.5 kcal/mol variance permitted for SAPT components should have allowed observation of correlation if it existed. The lack of correlation suggests the model parameter space is rugged. This in turn suggests orders of magnitude more QM dimer and cluster data would be needed to build a completely *ab initio* force field. This suspicion is given credence by the experience of the MB-pol and

GEM water models, both of which required thousands of structures to produce well-determined models.[96, 97]

***Limitations of a Classical Model.*** By nature of being a classical model, HIPPO has a set of limitations. As illustrated in the Results section, the agreement between experimental and *ab initio* data, while good, is not perfect. These inconsistencies generally arise because of the classical approximations the HIPPO model employs. In this section, we will briefly enumerate some of the most important limitations of the model. We will also rationalize why, despite these limitations, HIPPO is capable of agreement with experiment as good or better than some of the best published water models.

***Nuclear Quantum Effects.*** One of the most prominent areas for which the HIPPO water model is in disagreement with experiment is the heat capacity. This discrepancy is rooted in a physical effect that the HIPPO model does not directly address: Nuclear Quantum Effects or NQEs. NQEs show up in a variety of physical attributes of water. An instructive comparison is between $H_2O$ and $D_2O$, where $D_2O$ is meant to represent "classical" water with significantly less impact from nuclear quantum effects. The density of $D_2O$ is 0.3% smaller, the dielectric constant is 0.5% smaller, and the enthalpy of vaporization is 3.3% larger than those of $H_2O$. These are mostly small effects that have been largely accounted for *via* our parameterization procedure. The heat capacity, however, is different. $C_p$ at room temperature for $D_2O$ is 11% larger than that of $H_2O$. This difference is too large to be covered by flexibility in parameterization, and furthermore the nature of the difference makes it virtually impossible to do so with a classical model.

The root of all NQEs, but most especially the heat capacity effect, is the treatment of hydrogens as classical oscillators. According to the Born-Oppenheimer approximation, under which conventional molecular dynamics operates, both intra- and intermolecular vibrational modes of hydrogen in the HIPPO model are treated classically. This treatment is essentially incorrect from the standpoint of quantum mechanics, where the vibrations of hydrogen should be treated at quantum oscillators. The characteristic vibrations of a water molecule lie in the frequency range 1000-4000 $cm^{-1}$. However, at room temperature the amount of available thermal energy, $k_B T$, corresponds to a frequency of ~200 $cm^{-1}$. This means that for virtually all of the vibrational modes of hydrogen atoms in water, the spacing between energy levels is much greater than the amount of thermal energy available. At room temperature, corrections of 6 cal $mol^{-1}$ $K^{-1}$ and 2 cal $mol^{-1}$ $K^{-1}$ to account for this difference between quantum oscillators and the classical model have been proposed.[84, 98] Moreover, these considerations show that the magnitude of the error caused by imposing a classical model on a quantum system is temperature dependent. As temperature is decreased, vibrational excitation becomes more and more difficult as $k_B T$ drops. However, when temperature is increased, $k_B T$ becomes closer to the energy spacing of the hydrogen atom's low-frequency modes, allowing more vibrational excitation. The upshot of this temperature dependent error is that one should expect a classical water model to exhibit a heat of vaporization that is too high at low temperature and too low at high temperature. This is exactly the behavior seen in the HIPPO water model, giving it a heat capacity slightly higher than experiment.

Of course, the solution to fix this error in the heat capacity is to use a method that goes beyond the Born-Oppenheimer approximation to include NQEs. Other classical models have used methods such as path integral molecular dynamics (PIMD) or ring-polymer molecular dynamics (RPMD) with some success. Application of this methodology to the HIPPO model would be of great interest, given the otherwise high fidelity with experiment.

There are two likely reasons why HIPPO still attains good agreement with experiment despite not including nuclear quantum effects. The first is that for properties besides heat capacity, the impact of NQEs is small. The second reason is that while HIPPO is rooted in *ab initio* EDA calculations, it is not strictly an *ab initio* model. This means that there is some flexibility in parameterization that has allowed HIPPO to fit $H_2O$ experimental data without losing fidelity to the SAPT data from which it was originally derived. This flexibility is the driving force behind the parameterization process described in the methods section. In order to include NQEs implicitly, we optimized the initial *ab initio* derived parameters of the water model to reproduce $H_2O$ liquid properties.

***Many-Body Effects.*** The HIPPO model includes many-body effects through its polarization model. This induced dipole model allows for a linear order, classical electrostatic response of each atom to its environment. The results in the "Larger Clusters" section of the Results show that the model captures a majority of the total three-body energy of water clusters. However, there are other many-body effects which the HIPPO model does not include.

The first set of many-body effects excluded from HIPPO are "classical" electrostatic effects. These arise from terms involving higher-order polarizabilities, hyperpolarizabilities or charge transfer. Various water models such as NEMO and the ASP series of Stone and co-workers have included higher-order and hyperpolarizabilities.[99, 100] Similarly, there exist models for many-body charge transfer in water, such as those of Rick[101] as well as the forthcoming SIBFA water model.[102] The distinction between these various terms is not well defined and is presently the subject of intense scrutiny. However, their roots, regardless of nomenclature, are the same. They all describe the response of a molecule's electron density to its electrostatic environment to infinite order. HIPPO includes just the least computationally expensive leading term of the full expansion. Models that include higher-order or hyperpolarizabilities, or charge transfer are attempting to select those additional terms representing the largest additional portion of the full expansion. While HIPPO does include a pairwise charge-transfer term, the decision to not include any of the higher-order many-body effects derives from a simple observation. As shown in Figure 7 the missing part of the HIPPO three-body energy of water clusters is about 0.1 kcal/mol per molecule. This error is an order of magnitude smaller than other errors in the force field relative to *ab initio* results. The comparison indicates why HIPPO is capable of a high degree of agreement with experiment despite neglecting higher-order effects.

Of course, classical effects are not the only thing at play in intermolecular interactions. There exist many-body components to the dispersion and Pauli repulsion components of the intermolecular potential as well. There is a body of work showing that for some systems these quantum many-body effects, particularly many-body dispersion, can be important.[103-105] There are also a number of models available for including these effects in classical potentials.[106-108] However, the computational cost to include these effects for the purposes of the HIPPO model is prohibitive. Moreover, work from the Head-Gordon and co-workers has shown that the magnitude of these quantum many-body effects is insignificant for water-water and water-ion interactions.[72] Because many-body dispersion and repulsion account for less than 1% of the total many-body energy, even for close-contact water clusters, they are neglected by the current HIPPO model.

***How Good is SAPT for Water?*** A question one might ask is, "why not fit the HIPPO water model exclusively to SAPT data?" The suggestion in the question certainly has appeal. Fitting exclusively to SAPT would put HIPPO in the category of *"ab initio"* water models. The goal of the HIPPO project, however, is not to recreate any particular level of QM theory; it is to accurately predict experimental thermodynamic results. This is the reason we chose to refine the HIPPO parameters to reproduce the experimental density and heat of vaporization of liquid water. This strategy, however, leaves the HIPPO model in a middle ground that bears some explanation. Why use SAPT if the end result is ultimately fit to experiment?

To answer this question, it is helpful to look at the quality of an alternative "*ab initio"* version of the HIPPO water model. For the purposes of discussion, we will refer to this model as HIPPO-SAPT. Plotted in Figure 15 are the room temperature liquid properties of the stage one HIPPO-SAPT water model, fit to SAPT data, not yet refined for any experimental properties. This model is fit exclusively to SAPT2+ data on ~25 water dimer structures. Each of the components was fit individually, as outlined in the methods section. One can see from Figure 15 and Table 6 that the condensed phase properties of this model at room temperature are not far from the experimental values.

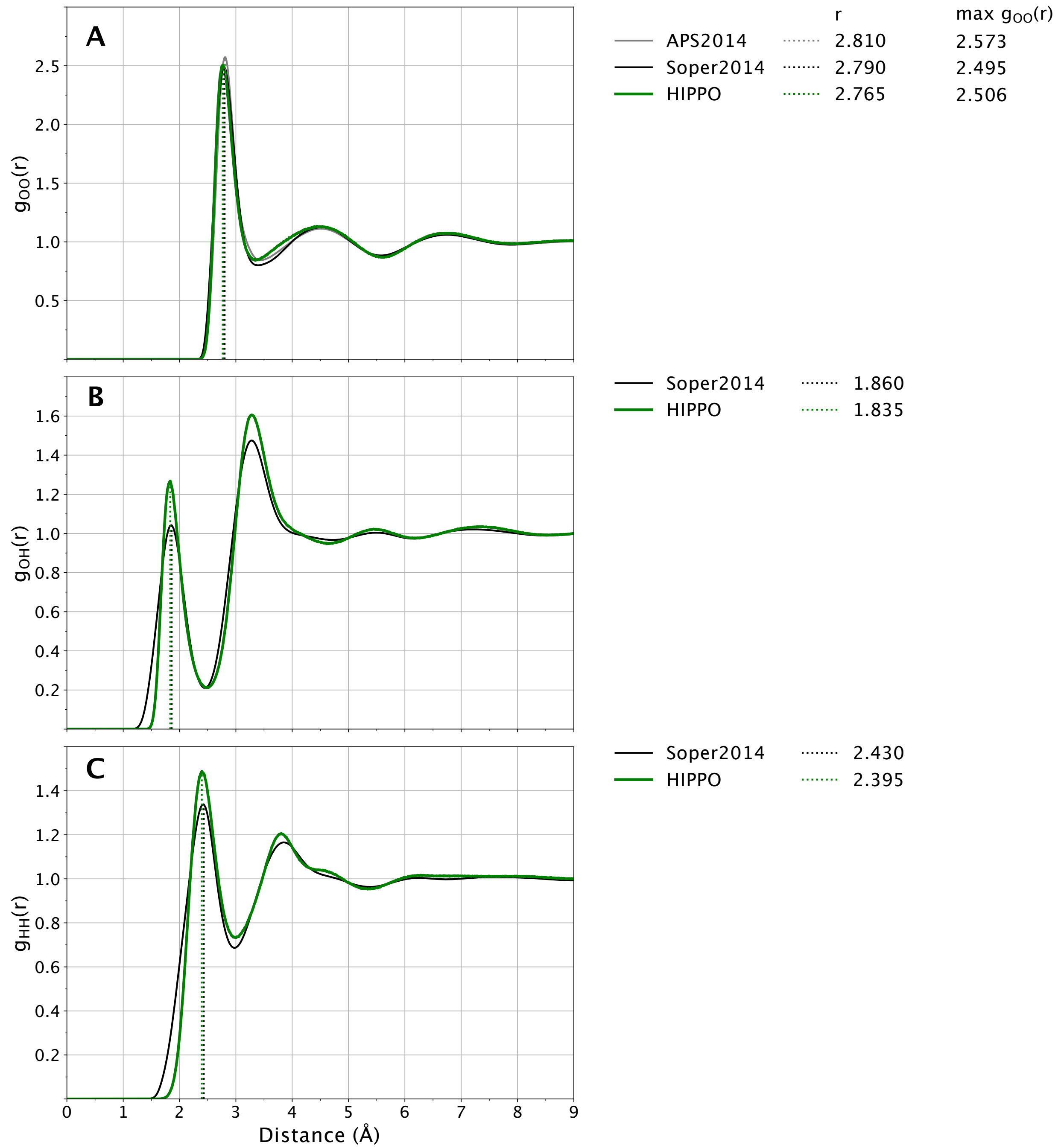

***Figure 15.*** *Water radial distribution function for the HIPPO-SAPT water model before optimization with ForceBalance at 298 K and 1 atm. HIPPO-SAPT results are shown in green and experimental in black. First peaks of the HIPPO-SAPT distribution are indicated with dotted green vertical lines. Experimental curves from references 81-83.*

***Table 6.*** *HIPPO-SAPT water properties at room temperature (298 K and 1 atm).*

| Property | HIPPO-SAPT | Experiment | Abs. Error (%) |
|---|---|---|---|
| Density (kg/m$^3$) | 974.195 | 997.045 | 22.85 (2.3) |
| Enthalpy of Vaporization (kJ/mol) | 41.194 | 43.989 | 2.795 (6.4) |
| Self-Diffusion Coefficient ($10^{-5}$ cm$^2$/s) | 2.805 | 2.230 | 0.575 (25.8) |
| Static Dielectric Constant | 80.050 | 78.409 | 1.641 (2.1) |

The radial distribution functions for O-O, O-H, and H-H are all in good agreement with experiment. The dielectric constant is also very close to the experimental value. The model is not perfect, however. There are significant discrepancies in the density, enthalpy of vaporization and self-diffusion coefficient, and the density vs. temperature curve for this model exhibits no maximum. These data indicate the quality of SAPT for water. Within the confines of the SAPT2+ level of theory, with an aug-cc-pVDZ basis set, SAPT is capable of producing a "rough" water model, but not one up to the accuracy of empirical polarizable force fields such as AMOEBA.

Because water is the most important component of any biomolecular force field, the level of accuracy of this SAPT "*ab initio*" force field is not sufficient. The general accuracy of the model from these initial parameters, however, tells us about the utility of using SAPT as a reference. HIPPO is built on a series of successive approximations. The model is fit to SAPT, but SAPT has some measurable error relative to CCSD(T), the so-called gold standard of quantum chemistry. CCSD(T), despite the title, however, is not perfect either. CCSD(T) uses the Born-Oppenheimer approximation, and as such is missing nuclear quantum effects (NQEs). This means that rather than using SAPT as a hard reference, the HIPPO strategy is instead to use SAPT as a guide. The SAPT data serves to solve the biggest problem in non-ab initio force fields: overdetermination. Requiring that HIPPO satisfy the SAPT components dramatically limits the parameter space available in the refinement phase of parameterization. This means that while it is not an *ab initio* force field, HIPPO is qualitatively different from empirical force fields because it follows the clearly identifiable series of approximations just described.

***Transferability.*** Within the confines of any particular functional form - density-based, point-charge or otherwise - there are an infinite number of equally good water models. This is a simple consequence of a problem that is overdetermined by its nature. Unlike the simplest fixed partial charge water models, many advanced or polarizable models have several tunable parameters, but a sparse number of experimental observables to fit against. What makes the HIPPO model unique is that it limits itself to a narrow window of parameter space by insisting that SAPT energy decomposition data be satisfied. This is true not just for water-water SAPT calculations. The final parameters of this water model produce an RMS error on the entire S101x7 database of less than 1.0 kcal/mol per component on average. This means that the relaxation of the parameters to fit liquid properties did not disrupt the backbone of the HIPPO framework. These data suggest that the HIPPO water model will not only reproduce pure water properties well, but also perform well as a solvent. Although yet untested, this natural fit between the water model and the rest of the future force field is important. Recent work has shown that various point charge water models can produce dramatically different results for protein simulations.[109] The emphasis on SAPT in the HIPPO model gives confidence that this water model will work well with the HIPPO small molecule and macromolecule models currently under development.

## Conclusions

The quality of a water model is a subjective quantity. The utility of a particular model depends upon the kinds of scientific questions one wants to answer. The "best" water model for a job will change depending on whether one wants a rough solvation model or a detailed comparison with spectroscopic values. For bulk phase properties, a number of water models traditionally used in molecular dynamics simulation, as well as more recent models, provide generally similar results. Importantly, however, models sufficiently accurate for homogeneous pure water simulation may not be appropriate to account for solvation by water in heterogeneous environments. Table 7 provides a minimal set of pure liquid properties for a subset of available models, including the HIPPO model described in this work.

***Table 7.*** *Selected properties of some water models used in MD simulation. HIPPO values are from the current work. Parameterization and data for other models are taken from: AMOEBA+,[53] AMOEBA03,[9] AMOEBA14,[110] TTM3-F,[111] SWM4-NDP and SWM6,[112] MB-POL,[62] MB-UCB,[113] TIP3P,[114] TIP4P-Ew[115] and TIP5P.[116] Dimer energy and heat of vaporization are in kcal/mol, density in g/cm$^3$, and diffusion coefficient as 10$^{-5}$ cm$^2$/s.*

| Model | $E_{dimer}$ | Density | $\Delta H_{vap}$ | Diffusion Coefficient | Dielectric Constant |
|---|---|---|---|---|---|
| Reference | -4.97 | 0.997 | 10.51 | 2.30 | 78.4 |
| HIPPO | -4.96 | 0.997 | 10.47 | 2.56 | 76.9 |
| AMOEBA+ | -4.85 | 0.998 | 10.6 | 2.14 | 78.8 |
| AMOEBA03 | -4.96 | 1.000 | 10.48 | 2.02 | 81 |
| AMOEBA14 | -4.64 | 0.998 | 10.63 | 2.36 | 79.4 |
| TTM3-F | -5.18 | 0.994 | 11.4 | 2.37 | 94.4 |
| SWM4-NDP | -5.15 | 0.994 | 10.45 | 2.85 | 78.0 |
| SWM6 | -5.27 | 0.996 | 10.52 | 2.14 | 78.1 |
| MB-POL | -5.05 | 1.007 | 10.93 | 2.8 | 68.4 |
| MB-UCB | -5.06 | 0.999 | 10.58 | – | – |
| TIP3P | -6.02 | 0.982 | 10.45 | 6.11 | 82 |
| TIP4P-Ew | -6.18 | 0.995 | 10.58 | 2.44 | 63.9 |
| TIP5P | -6.78 | 0.979 | 10.46 | 2.78 | 92 |

With the above in mind, we conclude by attempting to place the HIPPO water model in context. First, we lay out a general taxonomy of water models and attempt to place HIPPO in that scheme. Second, we present the level of accuracy one can expect when using a water model out of a particular class in the taxonomy. Third, we use these ideas to motivate exactly what the HIPPO model is intended to be useful for. And lastly, we summarize the main scientific points uncovered in the process of developing the HIPPO potential.

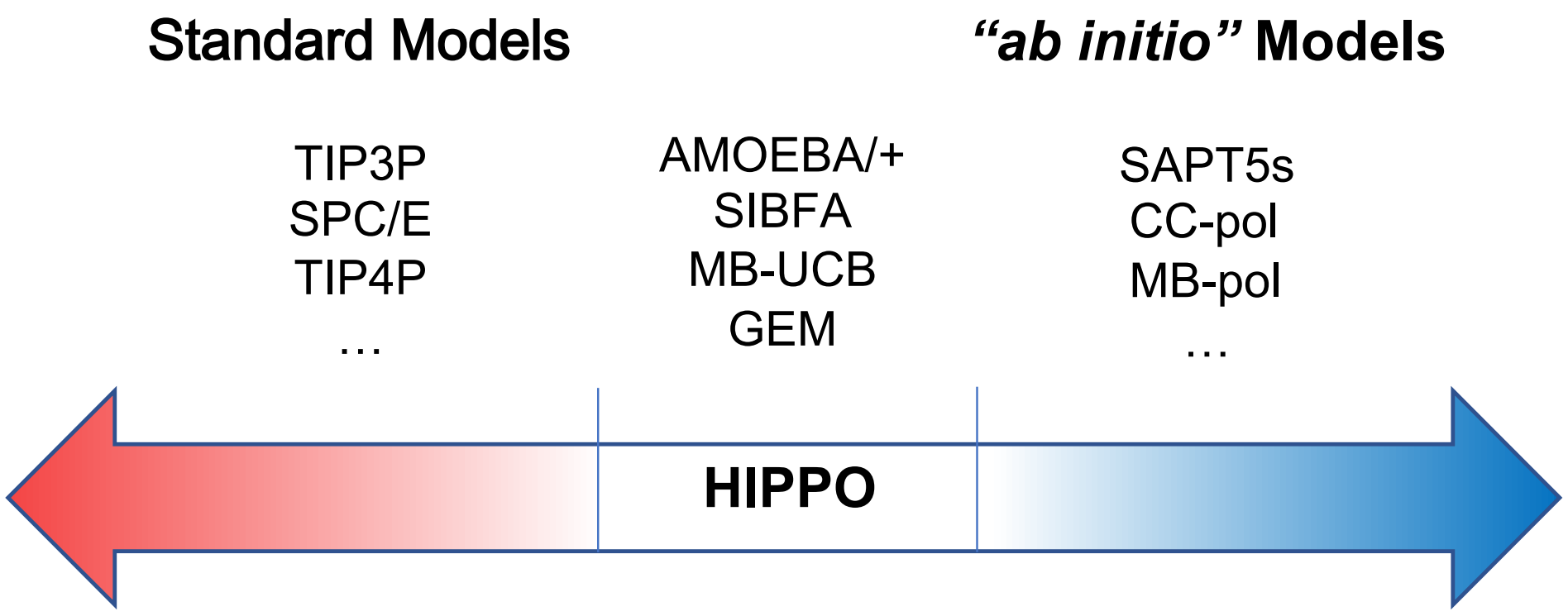

***Figure 16.*** *A non-exhaustive taxonomy of classical water models.*

Despite the staggering number of published classical water models, the existing atom-based models can be roughly grouped into three general categories: empirical, *ab initio* and physics-based. These three subsets loosely define a spectrum as illustrated in Figure 16, with empirical on one side, *ab initio* at the extreme, and physics-based in the middle. On the *ab initio* side are models fit solely to data from quantum mechanical calculations. On the other hand, empirical models are calibrated largely against experimental condensed phase properties. Models in the sparsely populated middle of the spectrum, which we term “physics-based”, attempt to reproduce both bulk phase and quantum mechanical calculation data simultaneously.

Examples of the empirical class of water models are the SPC and TIPS families of potentials. These may vary in the number and placement of interaction sites, but the functional form is essentially fixed: a Lennard-Jones van der Waals function coupled with point charge electrostatics. Because of this limited functional form, such models rely heavily on cancellation of errors. Thus, they are fit primarily to reproduce bulk phase properties of water around room temperature and pressure. The sheer number of published parameterizations of this functional form is a testament to how much flexibility is available during the fitting process. Because of this, most empirical models do give good agreement with the properties of water at room temperature, including a roughly correct description of the radial distribution function. However, these models are typically unable to capture fine-grained details of water structure. They struggle, for example, to correctly rank the ten Smith dimers or accurately predict the $2^{nd}$ virial coefficient. For this reason, parameterizations of general biomolecular force fields often are calibrated using a specific water model. The main advantage of balancing these model costs is speed. Empirical water models remain the tool of choice when extensive sampling or simulating large systems is of greater importance than quantitative model accuracy.

On the other end of the spectrum are the *ab initio* water models. These can be further subdivided into two camps: (1) unique models intended just for water, such as the ASP-W[117] and TTM[111] series, which rely on the unique electronic properties of water, and

(2) big data-derived water models, such as SAPT-5s,[118] CC-pol,[119] and MB-pol,[62] which are based on large amounts of high-quality *ab initio* data. What all *ab initio* models have in common, though, is they are primarily fit to reproduce quantum mechanical data. This gives them a level of accuracy much higher than empirical models. They generally give a high-fidelity description of the Born-Oppenheimer potential energy surface of water, and are able to capture the bulk property temperature dependence (modulo nuclear quantum effects) and detailed structural features of water very accurately. Moreover, many of these models are able to reproduce spectroscopic properties such as vibrational frequencies due to their fidelity to the underlying quantum mechanics. These qualities come with two major tradeoffs. First, because of their complexity, these models are generally much slower than empirical models. They are too slow, for instance, to efficiently sample biomolecular-sized systems. Second, the framework for these models is not easily generalizable to complex, heterogeneous systems. To date, none of the *ab initio* class of models have been successfully extended to produce a complete biomolecular force field.

The final class of water models lies in the space between empirical and *ab initio*. These "physics-based" models attempt to satisfy both quantum mechanical and bulk phase data simultaneously by employing more complex functional forms intended to directly approximate the underlying Born-Oppenheimer potential energy surface. Examples of models in this class are AMOEBA,[9, 110] AMOEBA+,[53] GEM,[16] SIBFA,[120] MB-UCB,[113] SWM,[112] and HIPPO. Because these models are classical approximations, the approximations used mean that there is a slight degradation of the Born-Oppenheimer surface compared to good *ab initio* models. For example, such physics-based models are generally not highly accurate for spectroscopic properties. Several of these models, however, are capable of quantitatively reproducing structural and energetic properties across a wider range of conditions. For instance, we have shown in this work that HIPPO is capable of quantitatively predicting water dimer properties, cluster energies, and the 2$^{nd}$ virial coefficient. Additionally, physics-based models, unlike the empirical class, can reliably reproduce the temperature dependence of bulk phase liquid properties. A detailed comparison of many-body energetics for the *ab Initio* MB-pol and TTM models against several physics-based polarizable water potentials was recently presented by Lambros and Paesani.[121] We agree with their conclusion that many-body *ab Initio* and physics-based water potentials should continue in parallel and with an eye toward ultimate convergence. HIPPO's use of the SAPT framework and explicit consideration of many-body energies within clusters is an initial step toward such convergence.

Having outlined what purposes best suit each class of model, the question is: If one needs a physics-based model, why consider HIPPO over the alternatives? For predicting many properties of water, HIPPO performs as well or better than the other listed models. However, this does not make HIPPO different in kind from the other models in its class. What makes HIPPO qualitatively different is the systematic, traceable series of approximations upon which it is constructed. HIPPO is based upon a model for charge density, from which every nonbonded term of the force field is derived. This allows the model to provide a direct approximation of Symmetry Adapted Perturbation Theory. SAPT is in turn approximate with respect to the current "gold standard" level of quantum chemistry, CCSD(T), which is in turn an approximation of the exact Born-Oppenheimer surface. Although the errors accumulated across this series of approximations place the derived HIPPO model too far from the exact potential surface to be a true *ab initio* model, this lineage gives HIPPO two properties that make it unique:

1. It dramatically limits the parameter search space for optimization against experimental data.
2. It gives a specific framework from which to build a more general, and complete molecular force field.

These qualities certainly contribute to the fidelity with which HIPPO predicts properties of water, but their primary value will lie in the ability to extend HIPPO to other molecular systems in the future.

An important point to make about the HIPPO force field is that despite its more complete and complex set of equations, the computational cost of the model is roughly equal to other physics-based models. In the Tinker9 and OpenMM 7.4.0 computer codes, both of which implement HIPPO and AMOEBA on GPUs, the difference in cost between the two models is negligible.

Several lessons were learned during the process of developing the HIPPO water model. First, SAPT2+ is insufficient to build an *ab initio* water model. Since the bulk properties of water are sensitive to small changes in the water dimer potential energy surface, we found that fitting only to SAPT2+ data could not produce a satisfactorily precise and accurate model. Second, the use of an underlying charge density is critical to the accurate modeling of both short- and long-range intermolecular interactions. This is obviously true for electrostatics, but it is no less true for other parts of the force field, including polarization, repulsion and dispersion. HIPPO shows that a charge density formulation can produce accurate many-body interactions vis-a-vis a polarization model, and we have demonstrated that a charge density model is also necessary to accurately reproduce van der Waals interactions. Third, atomic anisotropy is essential for a physics-based model, and is necessary to achieve fully correct behavior for water dimers and clusters. Importantly, we show this anisotropy is just as important in the repulsion component of the force field as in the traditional electrostatic portion. Furthermore, the HIPPO functional form illustrates that the nature of the anisotropy can be effectively captured by an energy model derived from the atomic multipole moments. Finally, and practically, we make the observation that dramatic improvement in the short-range physics of a force field can be incorporated without significant additional computational cost. Because the short-range terms have simple asymptotic behavior, the cost of HIPPO is comparable to or less than many of its physics-based force field peers.

The goal is for the HIPPO water model to become the cornerstone of a general force field for water, ions, organics and biomolecules. The critical importance of water as the solvent in many simulations justifies the high level of attention described in this work. The strength of interactions with monoatomic ions provides a useful stress test for new potentials. We are currently exploring HIPPO water-ion energetics along the lines of prior studies of AMOEBA water with ions.[122, 123] Continued parameterization for organic molecules and biomolecules will make use of the Caleman, *et. al* database of over 1200 experimental properties and values for 146 organic liquids,[124] and the S101x7 SAPT data set,[29] respectively. From the experience gained with water, the plan is to obtain atomic multipole values and polarizabilities from DMA and potential fitting.[10] Then we will use genetic and least squares optimization methods to fit liquid properties across multiple molecules simultaneously, using SAPT values from S101x7 data as guides via loose restraints. Lessons learned in the development of the HIPPO water model should prove useful as physics-based force fields progress toward maturity.

## Supporting Information

Supplementary tables including force field parameters for the HIPPO water model, comparison of Smith water dimer energies for published water models, and computed and experimental property values over the temperature and pressure ranges reported in the text (PDF).

## Acknowledgements

J.W.P. would like to thank the U.S. National Institutes of Health for support of the development of advanced force fields such as the HIPPO model and the associated Tinker molecular modeling software packages via Grant Nos. R01 GM106137 and R01 GM114237.

J.A.R. wishes to acknowledge funding from a Harry S. Truman Fellowship through the U.S. Department of Energy Laboratory Directed Research and Development program. Sandia National Laboratories is a multi-mission laboratory managed and operated by National Technology & Engineering Solutions of Sandia, LLC, a wholly owned subsidiary of Honeywell International Inc., for the U.S. Department of Energy's National Nuclear Security Administration under contract DE-NA0003525. This paper describes objective technical results and analysis. Any subjective views or opinions that might be expressed in the paper do not necessarily represent the views of the U.S. Department of Energy or the United States Government.

## Appendix A:

*Electrostatic Energy:*

*Core-Core:*

$$U_{core-core} = Z_i T_{ij} Z_j$$

$$T_{ij} = \frac{1}{R}$$

*Core-Density:*

$$U_{core-density} = Z_i \boldsymbol{T}^*_{\boldsymbol{ij}} \vec{M}_j$$

$$\boldsymbol{T}^*_{\boldsymbol{ij}} = [1 \quad \nabla \quad \nabla^2] T^*$$

$$T^* = \frac{1}{R} f_1^{damp}$$

$$\nabla T^* = -f_3^{damp} \frac{R_\alpha}{R^3}$$

$$\nabla^2 T^* = f_5^{damp} \frac{3R_\alpha R_\beta}{R^5} - f_3^{damp} \frac{\delta_{\alpha\beta}}{R^3}$$

$$f_1^{damp} = 1 - \left(1 + \frac{1}{2}\zeta_j R\right) e^{-\zeta_j R}$$

$$f_3^{damp} = 1 - \left(1 + \zeta_j R + \frac{1}{2}\left(\zeta_j R\right)^2\right) e^{-\zeta_j R}$$

$$f_5^{damp} = 1 - \left(1 + \zeta_j R + \frac{1}{2}\left(\zeta_j R\right)^2 + \frac{1}{6}\left(\zeta_j R\right)^3\right) e^{-\zeta_j R}$$

*Density-Density:*

$$U_{density-density} = \vec{M}_i \boldsymbol{T}_{\boldsymbol{ij}}^{\boldsymbol{overlap}} \vec{M}_j$$

$$\boldsymbol{T}_{\boldsymbol{ij}}^{\boldsymbol{overlap}} = \begin{bmatrix} 1 & \nabla & \nabla^2 \\ \nabla & \nabla^2 & \nabla^3 \\ \nabla^2 & \nabla^3 & \nabla^4 \end{bmatrix} \left(T^{overlap}\right)$$

$$T^{overlap} = \frac{1}{R} f_1^{overlap}$$

$$\nabla T^{overlap} = -f_3^{overlap} \frac{R_\alpha}{R^3}$$

$$\nabla^2 T^{overlap} = f_5^{overlap} \frac{3R_\alpha R_\beta}{R^5} - f_3^{overlap} \frac{\delta_{\alpha\beta}}{R^3}$$

$$\nabla^3 T^{overlap} = -f_7^{overlap}\frac{15R_\alpha R_\beta R_\gamma}{R^7} + f_5^{overlap}\frac{3\left(R_\alpha\delta_{\beta\gamma} + R_\beta\delta_{\alpha\gamma} + R_\gamma\delta_{\alpha\beta}\right)}{R^5}$$

$$\nabla^4 T^{overlap}$$

$$= f_9^{overlap}\frac{105R_\alpha R_\beta R_\gamma R_\eta}{R^9}$$

$$- f_7^{overlap}\frac{15\left(R_\alpha R_\beta\delta_{\gamma\eta} + R_\alpha R_\gamma\delta_{\beta\eta} + R_\alpha R_\eta\delta_{\beta\gamma} + R_\beta R_\gamma\delta_{\alpha\eta} + R_\beta R_\eta\delta_{\alpha\gamma} + R_\gamma R_\eta\delta_{\alpha\beta}\right)}{R^7}$$

$$+ f_5^{overlap}\frac{3\left(\delta_{\alpha\beta}\delta_{\gamma\eta} + \delta_{\alpha\gamma}\delta_{\beta\eta} + \delta_{\alpha\eta}\delta_{\beta\gamma}\right)}{R^5}$$

$$f_1^{overlap} = \begin{cases} 1 - \left(1 + \frac{11}{16}\zeta R + \frac{3}{16}(\zeta R)^2 + \frac{1}{48}(\zeta R)^3\right)e^{-\zeta R}, & \zeta_i = \zeta_j \\ 1 - A^2\left(1 + 2B + \frac{\zeta_i}{2}R\right)e^{-\zeta_i R} - B^2\left(1 + 2A + \frac{\zeta_j}{2}R\right)e^{-\zeta_j R}, & \zeta_i \neq \zeta_j \end{cases}$$

$$f_3^{overlap} = \begin{cases} 1 - \left(1 + \zeta R + \frac{1}{2}(\zeta R)^2 + \frac{7}{48}(\zeta R)^3 + \frac{1}{48}(\zeta R)^4\right)e^{-\zeta R}, & \zeta_i = \zeta_j \\ 1 - A^2\left(1 + \zeta_i R + \frac{1}{2}(\zeta_i R)^2\right)e^{-\zeta_i R} - B^2\left(1 + \zeta_j R + \frac{1}{2}\left(\zeta_j R\right)^2\right)e^{-\zeta_j R} - \\ \quad 2A^2B(1 + \zeta_i R)e^{-\zeta_i R} - 2B^2A\left(1 + \zeta_j R\right)e^{-\zeta_j R}, & \zeta_i \neq \zeta_j \end{cases}$$

$$f_5^{overlap} = \begin{cases} 1 - \left(1 + \zeta R + \frac{1}{2}(\zeta R)^2 + \frac{1}{6}(\zeta R)^3 + \frac{1}{24}(\zeta R)^4 + \frac{1}{144}(\zeta R)^5\right)e^{-\zeta R}, & \zeta_i = \zeta_j \\ 1 - A^2\left(1 + \zeta_i R + \frac{1}{2}(\zeta_i R)^2 + \frac{1}{6}(\zeta_i R)^3\right)e^{-\zeta_i R} - \\ B^2\left(1 + \zeta_j R + \frac{1}{2}\left(\zeta_j R\right)^2 + \frac{1}{6}\left(\zeta_j R\right)^3\right)e^{-\zeta_j R} - \\ 2A^2B\left(1 + \zeta_i R + \frac{1}{3}(\zeta_i R)^2\right)e^{-\zeta_i R} - \\ 2B^2A\left(1 + \zeta_j R + \frac{1}{3}\left(\zeta_j R\right)^2\right)e^{-\zeta_j R}, & \zeta_i \neq \zeta_j \end{cases}$$

$$
f_7^{overlap} = \begin{cases} 1 - \begin{pmatrix} 1 + \zeta R + \frac{1}{2}(\zeta R)^2 + \frac{1}{6}(\zeta R)^3 + \frac{1}{24}(\zeta R)^4 \\ + \frac{1}{120}(\zeta R)^5 + \frac{1}{720}(\zeta R)^6 \end{pmatrix} e^{-\zeta R}, \quad \zeta_i = \zeta_j \\ \\ 1 - A^2\left(1 + \zeta_i R + \frac{1}{2}(\zeta_i R)^2 + \frac{1}{6}(\zeta_i R)^3 + \frac{1}{30}(\zeta_i R)^4\right) e^{-\zeta_i R} - \\ B^2\left(1 + \zeta_j R + \frac{1}{2}(\zeta_j R)^2 + \frac{1}{6}(\zeta_j R)^3 + \frac{1}{30}(\zeta_j R)^4\right) e^{-\zeta_j R} - \\ 2A^2 B\left(1 + \zeta_i R + \frac{2}{5}(\zeta_i R)^2 + \frac{1}{15}(\zeta_i R)^3\right) e^{-\zeta_i R} - \\ 2B^2 A\left(1 + \zeta_j R + \frac{2}{5}(\zeta_j R)^2 + \frac{1}{15}(\zeta_j R)^3\right) e^{-\zeta_j R}, \qquad \zeta_i \neq \zeta_j \end{cases}
$$

$$
f_9^{overlap} = \begin{cases} 1 - \begin{pmatrix} 1 + \zeta R + \frac{1}{2}(\zeta R)^2 + \frac{1}{6}(\zeta R)^3 + \frac{1}{24}(\zeta R)^4 \\ + \frac{1}{120}(\zeta R)^5 + \frac{1}{720}(\zeta R)^6 + \frac{1}{5040}(\zeta R)^7 \end{pmatrix} e^{-\zeta R}, \quad \zeta_i = \zeta_j \\ \\ 1 - A^2\left(1 + \zeta_i R + \frac{1}{2}(\zeta_i R)^2 + \frac{1}{6}(\zeta_i R)^3 + \frac{4}{105}(\zeta_i R)^4 + \frac{1}{210}(\zeta_i R)^5\right) e^{-\zeta_i R} - \\ B^2\left(1 + \zeta_j R + \frac{1}{2}(\zeta_j R)^2 + \frac{1}{6}(\zeta_j R)^3 + \frac{4}{105}(\zeta_j R)^4 + \frac{1}{210}(\zeta_j R)^5\right) e^{-\zeta_j R} - \\ 2A^2 B\left(1 + \zeta_i R + \frac{3}{7}(\zeta_i R)^2 + \frac{2}{21}(\zeta_i R)^3 + \frac{1}{105}(\zeta_i R)^4\right) e^{-\zeta_i R} - \\ 2B^2 A\left(1 + \zeta_j R + \frac{3}{7}(\zeta_j R)^2 + \frac{2}{21}(\zeta_j R)^3 + \frac{1}{105}(\zeta_j R)^4\right) e^{-\zeta_j R}, \qquad \zeta_i \neq \zeta_j \end{cases}
$$

$$
B = \frac{\zeta_i^2}{\zeta_i^2 - \zeta_j^2}, \quad A = \frac{\zeta_j^2}{\zeta_j^2 - \zeta_i^2}
$$

# Appendix B:

*Permanent Electrostatic Field:*

(field at induced dipole i, due to permanent moments of atom j)

$$
\boldsymbol{F}_{\boldsymbol{i}}^{\boldsymbol{perm}}(R) = Z_j \nabla\left(\frac{1}{R}\right) + \mathrm{Q_j} \nabla\left(\frac{1}{R} f^{damp}(R)\right) + \boldsymbol{\mu}_{\mathbf{j}} \cdot \nabla^2\left(\frac{1}{R} f^{damp}(R)\right) + \boldsymbol{\Theta}_{\mathbf{j}} : \nabla^3\left(\frac{1}{R} f^{damp}(R)\right)
$$

$$
\nabla\left(\frac{1}{R} f^{damp}(R)\right) = -f_3^{damp} \frac{R_\alpha}{R^3}
$$

$$\nabla^2\left(\frac{1}{R}f^{damp}(R)\right) = f_5^{damp}\frac{3R_\alpha R_\beta}{R^5} - f_3^{damp}\frac{\delta_{\alpha\beta}}{R^3}$$

$$\nabla^3\left(\frac{1}{R}f^{damp}(R)\right) = -f_7^{damp}\frac{15R_\alpha R_\beta R_\gamma}{R^7} + f_5^{damp}\frac{3\left(R_\alpha\delta_{\beta\gamma} + R_\beta\delta_{\alpha\gamma} + R_\gamma\delta_{\alpha\beta}\right)}{R^5}$$

$$f_3^{damp} = 1 - \left(1 + \zeta_j R + \frac{1}{2}\left(\zeta_j R\right)^2\right)e^{-\zeta_j R}$$

$$f_5^{damp} = 1 - \left(1 + \zeta_j R + \frac{1}{2}\left(\zeta_j R\right)^2 + \frac{1}{6}\left(\zeta_j R\right)^3\right)e^{-\zeta_j R}$$

$$f_7^{damp} = 1 - \left(1 + \zeta_j R + \frac{1}{2}\left(\zeta_j R\right)^2 + \frac{1}{6}\left(\zeta_j R\right)^3 + \frac{1}{30}\left(\zeta_j R\right)^4\right)e^{-\zeta_j R}$$

*Induced Dipole Electrostatic Field:*

(field at induced dipole i, due to induced dipole j)

$$\boldsymbol{F_i^{ind}}(R) = \boldsymbol{\mu}_{\mathbf{j}}^{\boldsymbol{ind}} \cdot \nabla^2\left(\frac{1}{R}f^{overlap}(R)\right)$$

$$\nabla^2\left(\frac{1}{R}f^{overlap}(R)\right) = f_5^{overlap}\frac{3R_\alpha R_\beta}{R^5} - f_3^{overlap}\frac{\delta_{\alpha\beta}}{R^3}$$

$$f_3^{overlap} = \begin{cases} 1 - \left(1 + \zeta R + \frac{1}{2}(\zeta R)^2 + \frac{7}{48}(\zeta R)^3 + \frac{1}{48}(\zeta R)^4\right)e^{-\alpha R}, & \zeta_i = \zeta_j \\ 1 - A^2\left(1 + \zeta_i R + \frac{1}{2}(\zeta_i R)^2\right)e^{-\zeta_i R} - B^2\left(1 + \zeta_j R + \frac{1}{2}\left(\zeta_j R\right)^2\right)e^{-\zeta_j R} - \\ \quad 2A^2B(1 + \zeta_i R)e^{-\zeta_i R} - 2B^2A\left(1 + \zeta_j R\right)e^{-\zeta_j R}, & \zeta_i \neq \zeta_j \end{cases}$$

$$f_5^{overlap} = \begin{cases} 1 - \left(1 + \zeta R + \frac{1}{2}(\zeta R)^2 + \frac{1}{6}(\zeta R)^3 + \frac{1}{24}(\zeta R)^4 + \frac{1}{144}(\zeta R)^5\right)e^{-\zeta R}, & \zeta_i = \zeta_j \\ 1 - A^2\left(1 + \zeta_i R + \frac{1}{2}(\zeta_i R)^2 + \frac{1}{6}(\zeta_i R)^3\right)e^{-\zeta_i R} - \\ B^2\left(1 + \zeta_j R + \frac{1}{2}\left(\zeta_j R\right)^2 + \frac{1}{6}\left(\zeta_j R\right)^3\right)e^{-\zeta_j R} - \\ 2A^2B\left(1 + \zeta_i R + \frac{1}{3}(\zeta_i R)^2\right)e^{-\zeta_i R} - \\ 2B^2A\left(1 + \zeta_j R + \frac{1}{3}\left(\zeta_j R\right)^2\right)e^{-\zeta_j R}, & \zeta_i \neq \zeta_j \end{cases}$$

$$B = \frac{\zeta_i^2}{\zeta_i^2 - \zeta_j^2}, \quad A = \frac{\zeta_j^2}{\zeta_j^2 - \zeta_i^2}$$

## Appendix C:

*Pauli Repulsion:*

$$U_{ij} = \frac{K_i K_j}{R} S_{total}^2$$

$$\frac{S_{total}^2}{R} = \vec{M}_i \boldsymbol{T_{ij}^{repulsion}} \vec{M}_j$$

$$\boldsymbol{T_{ij}^{repulsion}} = \begin{bmatrix} 1 & \nabla & \nabla^2 \\ \nabla & \nabla^2 & \nabla^3 \\ \nabla^2 & \nabla^3 & \nabla^4 \end{bmatrix} \left(T^{pauli}\right)$$

$$T^{pauli} = \frac{\zeta_i^3 \zeta_j^3}{R} f_1^{rep}$$

$$\nabla T^{pauli} = -f_3^{rep} R_\alpha$$

$$\nabla^2 T^{pauli} = f_5^{rep} 3 R_\alpha R_\beta - f_3^{rep} \delta_{\alpha\beta}$$

$$\nabla^3 T^{pauli} = -f_7^{rep} 15 R_\alpha R_\beta R_\gamma + f_5^{rep} 3\left(R_\alpha \delta_{\beta\gamma} + R_\beta \delta_{\alpha\gamma} + R_\gamma \delta_{\alpha\beta}\right)$$

$$\begin{aligned} \nabla^4 T^{pauli} = {} & f_9^{rep} 105 R_\alpha R_\beta R_\gamma R_\eta \\ & - f_7^{rep} 15\left(R_\alpha R_\beta \delta_{\gamma\eta} + R_\alpha R_\gamma \delta_{\beta\eta} + R_\alpha R_\eta \delta_{\beta\gamma} + R_\beta R_\gamma \delta_{\alpha\eta} + R_\beta R_\eta \delta_{\alpha\gamma} \right. \\ & \left. + R_\gamma R_\eta \delta_{\alpha\beta}\right) + f_5^{rep} 3\left(\delta_{\alpha\beta}\delta_{\gamma\eta} + \delta_{\alpha\gamma}\delta_{\beta\eta} + \delta_{\alpha\eta}\delta_{\beta\gamma}\right) \end{aligned}$$

$$f_1^{rep} = \left(f_{exp}\right)^2$$

$$f_{exp} = \begin{cases} \frac{1}{\zeta^3}\left(1 + \frac{\zeta R}{2} + \frac{1}{3}\left(\frac{\zeta R}{2}\right)^2\right) e^{\frac{-\zeta R}{2}}, & \zeta_i = \zeta_j \\ \frac{1}{2X^3 R}\left[\zeta_i\left(RX - 2\zeta_j\right) e^{\frac{-\zeta_j R}{2}} + \zeta_j\left(RX + 2\zeta_i\right) e^{\frac{-\zeta_i R}{2}}\right], & \zeta_i \neq \zeta_j \end{cases},$$

$$X = \left(\frac{\zeta_i}{2}\right)^2 - \left(\frac{\zeta_j}{2}\right)^2 .$$

$$f_3^{rep} = 2 f_{exp} f_{exp}'$$

$$f_{exp}' = \begin{cases} \frac{1}{\zeta^3}\frac{1}{3}\left(\frac{\zeta}{2}\right)^2\left(1 + \frac{\zeta R}{2}\right) e^{\frac{-\zeta R}{2}}, & \zeta_i = \zeta_j \\ \frac{1}{2X^3 R}\left[\left(\frac{1}{2}\zeta_i\zeta_j X - \frac{\zeta_i \zeta_j^2}{R} - \frac{2\zeta_i\zeta_j}{R^2}\right) e^{\frac{-\zeta_j R}{2}} + \left(\frac{1}{2}\zeta_i\zeta_j X + \frac{\zeta_j \zeta_i^2}{R} + \frac{2\zeta_i\zeta_j}{R^2}\right) e^{\frac{-\zeta_i R}{2}}\right], & \zeta_i \neq \zeta_j \end{cases}$$

$$f_5^{rep} = 2\left(f_{exp} f_{exp}'' + f_{exp}' f_{exp}'\right)$$

$$f''_{exp} = \begin{cases} \dfrac{1}{\zeta^3}\dfrac{1}{9}\left(\dfrac{\zeta}{2}\right)^4 e^{\frac{-\zeta R}{2}}, & \zeta_i = \zeta_j \\ \dfrac{1}{2X^3R^2}\left[\begin{array}{l} \left(\dfrac{1}{4}\zeta_i\zeta_j^2 X - \dfrac{\zeta_i\zeta_j^3}{2R} + \dfrac{\zeta_i\zeta_j X}{2R} - \dfrac{3\zeta_i\zeta_j^2}{R^2} - \dfrac{6\zeta_i\zeta_j}{R^5}\right)e^{\frac{-\zeta_j R}{2}} + \\ \left(\dfrac{1}{4}\zeta_j\zeta_i^2 X + \dfrac{\zeta_j\zeta_i^3}{2R} + \dfrac{\zeta_j\zeta_i X}{2R} + \dfrac{3\zeta_j\zeta_i^2}{R^2} + \dfrac{6\zeta_j\zeta_i}{R^5}\right)e^{\frac{-\zeta_i R}{2}} \end{array}\right], & \zeta_i \neq \zeta_j \end{cases}$$

$$f_7^{rep} = 2\left(f_{exp}f'''_{exp} + 3f''_{exp}f'_{exp}\right)$$

$$f'''_{exp} = \begin{cases} \dfrac{1}{\zeta^3}\dfrac{1}{45}\left(\dfrac{\zeta}{2}\right)^5 \dfrac{1}{R} e^{\frac{-\zeta R}{2}}, & \zeta_i = \zeta_j \\ \dfrac{1}{2X^3R^3}\left[\begin{array}{l} \left(\dfrac{1}{8}\zeta_i\zeta_j^3 X + \dfrac{3}{4}\dfrac{\zeta_i\zeta_j^2 X}{R} + \dfrac{3}{2}\dfrac{\zeta_i\zeta_j X}{R^2} - \dfrac{1}{4}\dfrac{\zeta_i\zeta_j^4}{R} - \dfrac{3\zeta_i\zeta_j^3}{R^2} - \dfrac{15\zeta_i\zeta_j^2}{R^3} - \dfrac{30\zeta_i\zeta_j}{R^4}\right)e^{\frac{-\zeta_j R}{2}} + \\ \left(\dfrac{1}{8}\zeta_j\zeta_i^3 X + \dfrac{3}{4}\dfrac{\zeta_j\zeta_i^2 X}{R} + \dfrac{3}{2}\dfrac{\zeta_j\zeta_i X}{R^2} + \dfrac{1}{4}\dfrac{\zeta_j\zeta_i^4}{R} + \dfrac{3\zeta_j\zeta_i^3}{R^2} + \dfrac{15\zeta_j\zeta_i^2}{R^3} + \dfrac{30\zeta_j\zeta_i}{R^4}\right)e^{\frac{-\zeta_i R}{2}} \end{array}\right], & \zeta_i \neq \zeta_j \end{cases}$$

$$f_9^{rep} = 2\left(f_{exp}f''''_{exp} + 4f'''_{exp}f'_{exp} + 3f''_{exp}f''_{exp}\right)$$

$$f''''_{exp} = \begin{cases} \dfrac{1}{\zeta^3}\dfrac{1}{315}\left(\dfrac{\zeta}{2}\right)^5 \dfrac{1}{R^3}\left(1 + \dfrac{\zeta R}{2}\right) e^{\frac{-\zeta R}{2}}, & \zeta_i = \zeta_j \\ \dfrac{1}{2X^3R^4}\left[\begin{array}{l} \left(\begin{array}{l} \dfrac{1}{16}\zeta_i\zeta_j^4 X + \dfrac{3}{4}\dfrac{\zeta_i\zeta_j^3 X}{R} + \dfrac{15}{4}\dfrac{\zeta_i\zeta_j^2 X}{R^2} + \dfrac{15}{2}\dfrac{\zeta_i\zeta_j X}{R^3} - \\ \dfrac{1}{8}\dfrac{\zeta_i\zeta_j^5}{R} - \dfrac{5}{2}\dfrac{\zeta_i\zeta_j^4}{R^2} - \dfrac{45}{2}\dfrac{\zeta_i\zeta_j^3}{R^3} - \dfrac{105\zeta_i\zeta_j^2}{R^4} - \dfrac{210\zeta_i\zeta_j}{R^5} \end{array}\right)e^{\frac{-\zeta_j R}{2}} + \\ \left(\begin{array}{l} \dfrac{1}{16}\zeta_j\zeta_i^4 X + \dfrac{3}{4}\dfrac{\zeta_j\zeta_i^3 X}{R} + \dfrac{15}{4}\dfrac{\zeta_j\zeta_i^2 X}{R^2} + \dfrac{15}{2}\dfrac{\zeta_j\zeta_i X}{R^3} + \\ \dfrac{1}{8}\dfrac{\zeta_j\zeta_i^5}{R} + \dfrac{5}{2}\dfrac{\zeta_j\zeta_i^4}{R^2} + \dfrac{45}{2}\dfrac{\zeta_j\zeta_i^3}{R^3} + \dfrac{105\zeta_j\zeta_i^2}{R^4} + \dfrac{210\zeta_j\zeta_i}{R^5} \end{array}\right)e^{\frac{-\zeta_i R}{2}} \end{array}\right], & \zeta_i \neq \zeta_j \end{cases}$$